\documentclass[superscriptaddress,aps,10pt,twocolumn,prb,
nofootinbib]{revtex4}

\usepackage{amssymb}
\usepackage{amsmath}
\usepackage{amsfonts}
\usepackage{graphicx}
\usepackage{xcolor}
\usepackage{soul}
\usepackage{float}
\usepackage{diagbox}

\newcommand{\bfmath}[1]{\boldsymbol{#1}}

\newcommand{\grad}{\bfmath{\nabla}}
\setcitestyle{numbers,square,citesep={,\kern-.3em}}
\begin{document}

\title{Eigenenergies of excitonic giant-dipole states in cuprous oxide}

\date{\today}
\author{Markus Kurz}
\email{markus.kurz@uni-rostock.de}
\author{Stefan Scheel}
\affiliation{Institut f\"ur Physik, Universit\"at Rostock,  
Albert-Einstein-Stra{\ss}e 23, 18059 Rostock, Germany}
\begin{abstract}
In this work we present the eigenspectra of a novel species of Wannier excitons 
when exposed to crossed electric and magnetic fields. In particular, we compute
the eigenenergies of giant-dipole excitons in $\textrm{Cu}_2\textrm{O}$ in 
crossed fields. In our theoretical approach, we calculate the excitonic spectra 
within both an approximate as well as a numerically exact approach for arbitrary 
field configurations. We verify that stable bound excitonic giant-dipole states 
are only possible in the strong magnetic field limit, as this is the only 
regime providing sufficiently deep potential wells for their existence. 
Comparing both analytic as well as numerical calculations, we obtain excitonic 
giant-dipole spectra with level spacings in the range of $0.6\ldots100\, \mu 
\textrm{eV}$.
\end{abstract}
\maketitle
\section{Introduction}
In a semiconductor environment, excitons are the quanta of the fundamental 
optical excitation which consist of a negatively charged electron in the 
conduction band and a positively charged hole in the valence band 
\cite{Frenkel1931, Mott1938}.  
As the interaction between them can be modeled as a screened Coulomb 
interaction, excitons are often considered to be a solid-state quasi-particle 
analogue to the hydrogen atom \cite{Gross1952,Hayashi1952,Gross1956}. In recent 
times, the measurement of hydrogen-like absorption spectrum of these 
quasi-particles up to principal quantum numbers of $n=25$ in cuprous oxide 
($\textrm{Cu}_2\textrm{O}$) have attracted attention 
\cite{Scheel2014}.
However, the hydrogen-like model of excitons is generally too simplistic, and
has been expanded by taking into account the complex valence band structure and 
the cubic symmetry $O_{\rm h}$ of $\textrm{Cu}_2\textrm{O}$ 
\cite{Lipari1970,Lipari1973a,Lipari1973b,Uihlein1981,Suzuki1974,Schoene2016}. 
This ansatz has been both theoretically and experimentally successfully applied 
for describing the correct level structure due to fine- and hyperfine splitting 
of excitonic states \cite{Thewes2015}. 

The addition of external electric and magnetic fields further reduces the 
symmetry of the exciton states, thereby leading to level structures 
possessing numerous complex splittings of excitonic absorption lines 
\cite{Agekyan1977,Altarelli1972,Cho1974}. For instance, high-resolution 
transmission spectroscopy of excitons in cuprous oxide subject to an external 
electric field increases the complexity of the measured spectra with increasing 
field strength. In particular, excitonic states with different parity become 
mixed, leading to optical activation of states which remain dark in zero 
external field \cite{Semina2016a,Semina2016b}.
Furthermore, recent high-resolution spectroscopy and theoretical modeling
of excitons in $\textrm{Cu}_2\textrm{O}$ have provided a fundamental 
understanding of complex absorption spectra in external magnetic fields for 
field strengths of up to $7\, \rm T$ and excitonic states with principal 
quantum numbers $n \le 7$ \cite{Schweiner2016,Schweiner2016b}. As the cubic 
lattice and the external magnetic field break all anti-unitary symmetries, 
several studies have shown that magneto-excitons in $\textrm{Cu}_2\textrm{O}$ 
obey GUE (Gaussian unitary ensemble) statistics 
\cite{Schweiner2017,Schweiner2017b,Schweiner2017c}.

In the case of field-dressed excitonic species, the total momentum of the 
system is not conserved, and an exact separation of the relative and 
center-of-mass degrees of freedom is impossible \cite{Schmelcher1993a}. There 
exists, however, an alternative conserved quantity, the so-called 
pseudomomentum, with whose help one can carry out a pseudoseparation of the 
center-of-mass and relative motion for neutral systems. In a recently article, 
a theoretical description of field-dressed excitons in 
$\textrm{Cu}_2\textrm{O}$ 
has been developed \cite{Kurz2017}. There, it has been shown that the effect of 
the center-of-mass degrees of freedom on the internal motion is an effective 
potential that gives rise to a number of outer potential wells for certain 
values of the pseudomomentum and applied field strengths. Potentially bound 
states in these outer potential wells are of decentered character with an 
electron-hole separation of up to several micrometers, leading to huge 
permanent electric dipole moments, thereby justifying the label excitonic 
giant-dipole states. Its counterpart in atomic physics, i.e. atomic 
giant-dipole states, have been predicted theoretically \cite{Baye1992, 
Dzyaloshinskii1992,Schmelcher1993a,Dippel1994, Schmelcher1998,Schmelcher2001} 
and explored experimentally in the early 1990's \cite{Fauth1987,Raithel1993}. 

Although the first study on excitonic giant-dipole potential surfaces has 
provided strong indications for the existence of excitonic giant-dipole states, 
a systematic analysis of their bound-state properties, such as binding energies 
and energy spectra, is still missing. In this work, we extend previous studies 
by deriving the irreducible tensor representation of field-dressed excitons, and
calculating the eigenenergies of giant-dipole states in 
$\textrm{Cu}_2\textrm{O}$. Here, we employ both approximate as well as 
numerically exact approaches.

This paper is organized as follows. In Sec.~\ref{gd_Hamiltonian}, we present 
the Hamiltonian of excitons in crossed electric and magnetic fields in its 
irreducible representation. Following this, in Sec.~\ref{strong_e_field}, we 
analyze the possibility of bound excitonic giant-dipole states in the limit of 
strong electric fields. Within this regime, we perform an adiabatic 
approximation that provides us with the possibility to derive analytic results. 
We find that, in this limiting regime, no bound states are present due to 
insufficiently deep potential energy surfaces. Following the adiabatic 
approach, we perform a similar analysis for arbitrary electric and magnetic 
field strengths in Sec.~\ref{arbitrary_fields}. We find that, in the strong 
magnetic field limit, the potential surfaces are sufficiently deep to provide 
bound states within the local potential minima. In Sec.~\ref{full_spectra}, we 
finally consider full couplings between the potential surfaces and calculate 
the excitonic eigenspectra within an exact diagonalization approach for various 
field strengths and field orientations.  

\section{The excitonic giant-dipole Hamiltonian}
\label{gd_Hamiltonian}
The Wannier excitons in $\textrm{Cu}_2\textrm{O}$ analyzed in this work 
are formed by an electron in the lowest $\Gamma^{+}_{6}$-conduction band and a 
positively charged hole in the uppermost (triply degenerate) 
$\Gamma^{+}_{5}$-valence band. The energy gap between the two bands is
$E_{g}=2.17208\, \textrm{eV}$ \cite{Scheel2014}. In contrast to the conduction 
band, the three uppermost valence bands are deformed due to interband 
interactions and the non-spherical symmetry of the crystal. These properties 
can be represented by an effective $I=1$ quasi-spin representation in the hole 
degrees of freedom \cite{Suzuki1974}.

In crossed electric and magnetic fields, the excitonic system possesses a 
constant of motion, the so-called pseudomomentum $\hat{\bfmath{K}}$ with
\begin{eqnarray}
\hat{\bfmath{K}}=\bfmath{P}-\frac{1}{2}\bfmath{B} \times \bfmath{r},\ \ \ \bfmath{r}=\bfmath{r}_e-\bfmath{r}_h,
\end{eqnarray}
and eigenvalues $\bfmath{K}$ \cite{Avron1978,Herold1981,Johnson1983}. As it has 
been discussed in detail previously, the excitonic Hamiltonian $H_{\rm ex}$ can 
be transformed into a single-particle Hamiltonian \cite{Schweiner2016b,Kurz2017}
\begin{equation}
H_{\rm ex}=H_0+H_{\rm so}+H_{\rm B},
\label{Hex}
\end{equation}
with
\begin{eqnarray}
&\ &H_0 =\frac{\bfmath{\pi}^2}{2m_e}+H_{h}(\bfmath{\pi})+V(\bfmath{r}),\ \ H_{\rm so}=\frac{2}{3}\bar{\Delta}(1+\bfmath{I}\cdot\bfmath{S}_h), 
\nonumber\\
&\ &H_{\rm B}=\bar{\mu}_B [(3\kappa+\frac{g_s}{2})\bfmath{I}\cdot
\bfmath{B}-g_s\bfmath{S}_{h}\cdot\bfmath{B}].
\label{Hexsingle}
\end{eqnarray}
The first term in $H_0$ stems from the kinetic energy of the electron whose 
effective mass $m_e=0.985m_0$ is almost identical to the free electron mass 
$m_0$. The second term is the hole Hamiltonian 
\begin{eqnarray}
H_{h}(\bfmath{\pi})&=&\frac{\bfmath{\pi}^{2}}{2m_0}(\gamma_1+4\gamma_2)
-\frac{3\gamma_2}{m_0}(\{ \pi^{2}_{x}I^{2}_{x}\}+{\rm c.p.})\nonumber\\
&\ &-\frac{6\gamma_3}{m_0}[\{\{\pi_{x}\pi_{y}\}\{I_xI_y\}\}+{\rm c.p.}]
\label{Ham_hole}
\end{eqnarray}
which is more complex due to the three coupled valence bands. The material 
parameters $\gamma_{i},\ i=1,2,3$ are the so-called Luttinger parameter and 
characterize the considered material \cite{Luttinger1954,Luttinger1956}. The 
values for $\textrm{Cu}_2\textrm{O}$ are given in 
Appendix~\ref{app_exc_paramteter}. The mapping $\{a b\} = (ab + ba)/2$ is the 
symmetric product and c.p. denotes cyclic permutations \cite{Suzuki1974}. 

The term $H_{\rm so}$ denotes the spin-orbit coupling of the hole-spin 
$\bfmath{S}_h$ with $\bfmath{I}$, while $H_B$ includes the coupling of the hole 
spins to the external magnetic field. As we do not include any kind of 
electronic spin-orbit coupling or spin-spin interaction, the electron spin 
$\bfmath{S}_e$ is not considered throughout this work. If not stated otherwise, 
we use excitonic Hartree units throughout this work, i.e.\ 
$e=\hbar=m_0/\gamma^{\prime}=1/4\pi \varepsilon_0 \varepsilon=1$ (see 
Appendix~\ref{app_exc_paramteter}). Here, $\varepsilon = 7.5$ is the
static dielectric constant of the bulk material and 
$\gamma^{\prime}_{1} \equiv m_0/m_e + \gamma_1$.

The quantity $\bfmath{\pi}$ is a generalized kinetic momentum which contains, 
besides the configuration space degrees of freedom $\bfmath{p}$ and $\bfmath{r}$, the spin-1 matrices $I_i,\ 
i=1,2,3$. In an arbitrary gauge, its components $\pi_i$ are given by 
\cite{Kurz2017}
\begin{eqnarray}
\pi_i=1_Ip_i-q A_i(\bfmath{r})+\partial_i f-\sum_{k}
(\frac{m_h}{M}1_I\delta_{ki}-\Omega_{ki})\tilde{K}_k\label{pi}
\end{eqnarray}
with $M=m_e+m_h$ and 
\begin{eqnarray*}
q=\frac{m_e-m_h}{M},\ \bfmath{A}(\bfmath{r})=\frac{1}{2}\bfmath{B}\times 
\bfmath{r},\ \tilde{\bfmath{K}}=\bfmath{K}+\bfmath{B}\times \bfmath{r},
\end{eqnarray*}
where $m_h \equiv m_0/\gamma_1$ denotes the hole mass. As the function $f$ can 
be eliminated via a simple gauge transformation, it will no longer be
considered. Together with the first term in Eq.~(\ref{Hexsingle}), one can 
define a kinetic energy Hamiltonian 
\begin{eqnarray}
T(\bfmath{\pi})=\frac{\bfmath{\pi}^2}{2m_e}+ H_{h}(\bfmath{\pi})
\end{eqnarray}
which parametrically depends on the pseudomomentum $\bfmath{K}$. The last term 
in $H_0$ represents a potential term $V$ that reads as
\begin{eqnarray}
\hspace{-0.5cm}V(\bfmath{r})&=&\left(\Omega_1 
\tilde{K}^2+\bfmath{E}\cdot\bfmath{r}-\frac{1}{r}\right) 
1_I-\Omega_2\sum_i\tilde{K}^{2}_{i}I_{ii}\nonumber\\ &\ 
&-\frac{2}{3}\Omega_3\sum_{ij,j<i}\tilde{K}_{i}\tilde{K}_{j}I_{ij},\ 
\tilde{\bfmath{K}}=\bfmath{K}+\bfmath{B}\times \bfmath{r}.\label{Vgd}
\end{eqnarray}
It describes an effective two-body potential including the electron-hole 
Coulomb interaction, the Stark coupling, and magnetic field terms. Together 
with $H_{\rm so}$ and $H_{\rm B}$, it defines the exact electron-hole 
potential 
\begin{eqnarray}
V_{\rm gd}(\bfmath{r})=V(\bfmath{r})+ H_{\rm so}+H_{\rm B}
\end{eqnarray}
for field-dressed excitons in cuprous oxide \cite{Kurz2017}. 

Using the vector components $\pi_i$ and $\tilde{K}_i$, one can define the
symmetric and trace-free Cartesian tensor operators
\begin{eqnarray}
I_{ij}&=&3\{ I_i I_j \}- 2\delta_{ij},\ \ 
\Pi_{ij}=3\{ \pi_i \pi_j \}- \bfmath{\pi}^2\delta_{ij},\\
\tilde{K}_{ij}&=&3\tilde{K}_i\tilde{K}_j-\tilde{K}^2\delta_{ij},\\
\Xi_{ij}&=&\left( \Pi_{ij}+m_0(\frac{\Omega_2}{\gamma_2}\delta_{ij}
+\frac{\Omega_3}{3\gamma_3}(1-\delta_{ij}))\tilde{K}_{ij}\right).
\end{eqnarray}
Using these tensor operators we derive the irreducible representation of 
the excitonic Hamiltonian given by Eq.~(\ref{Hex}), and we obtain
\begin{eqnarray}
&\ &\hspace{-0.5cm}H_{\rm ex}=\frac{\bfmath{\pi}^2}{2}-\frac{\mu'}{3} \{ \Xi^{(2)} \cdot I^{(2)} 
\} +\frac{\delta'}{3}\big( \sum_{k = \pm 4} \{ [\Xi^{(2)}\times 
I^{(2)}]^{(4)}_{k} \}\nonumber\\
&\ &\hspace{-0.5cm}+ \frac{\sqrt{70}}{5} \{ [\Xi^{(2)} \times 
I^{(2)}]^{(4)}_{0} \} \big)+\left(\Omega_1 \tilde{K}^2+E^{(1)} \cdot r^{(1)}-\frac{1}{r}\right) 
\nonumber\\
&\ &\hspace{-0.5cm}+H_{\rm so}+H_{\rm B} \label{Hex_irr}
\end{eqnarray}
with $\mu'=(3\gamma_3+2\gamma_2)/5\gamma^{'}_{1}$ and 
$\delta^{'}=(\gamma_3-\gamma_2)/2\gamma^{'}_{1}$. 
The mapping
\begin{gather}
\{ [\Xi^{(2)} \times I^{(2)}]^{(4)}_{k} \} \nonumber\\
\equiv \frac{1}{2}\left( [\Xi^{(2)} 
\times I^{(2)}]^{(4)}_{k} + [I^{(2)} \times \Xi^{(2)}]^{(4)}_{k} \right) 
\end{gather}
reflects the fact that the Cartesian tensor components $\Xi_{ij}$ and $I_{kl}$ 
do not necessarily commute. We note that this Hamiltonian is the most compact 
irreducible tensor representation of excitons in external electric and magnetic 
fields for arbitrary field strengths and field directions. Obviously, one can 
derive irreducible representations for kinetic and potential energy terms 
separately. These can be found in Appendix~\ref{t_v_irr_rep}. 

\section{Adiabatic approximation}
\label{adiabatic_approx}
\begin{figure}[ht]
\centering
\includegraphics[width=0.475\textwidth]{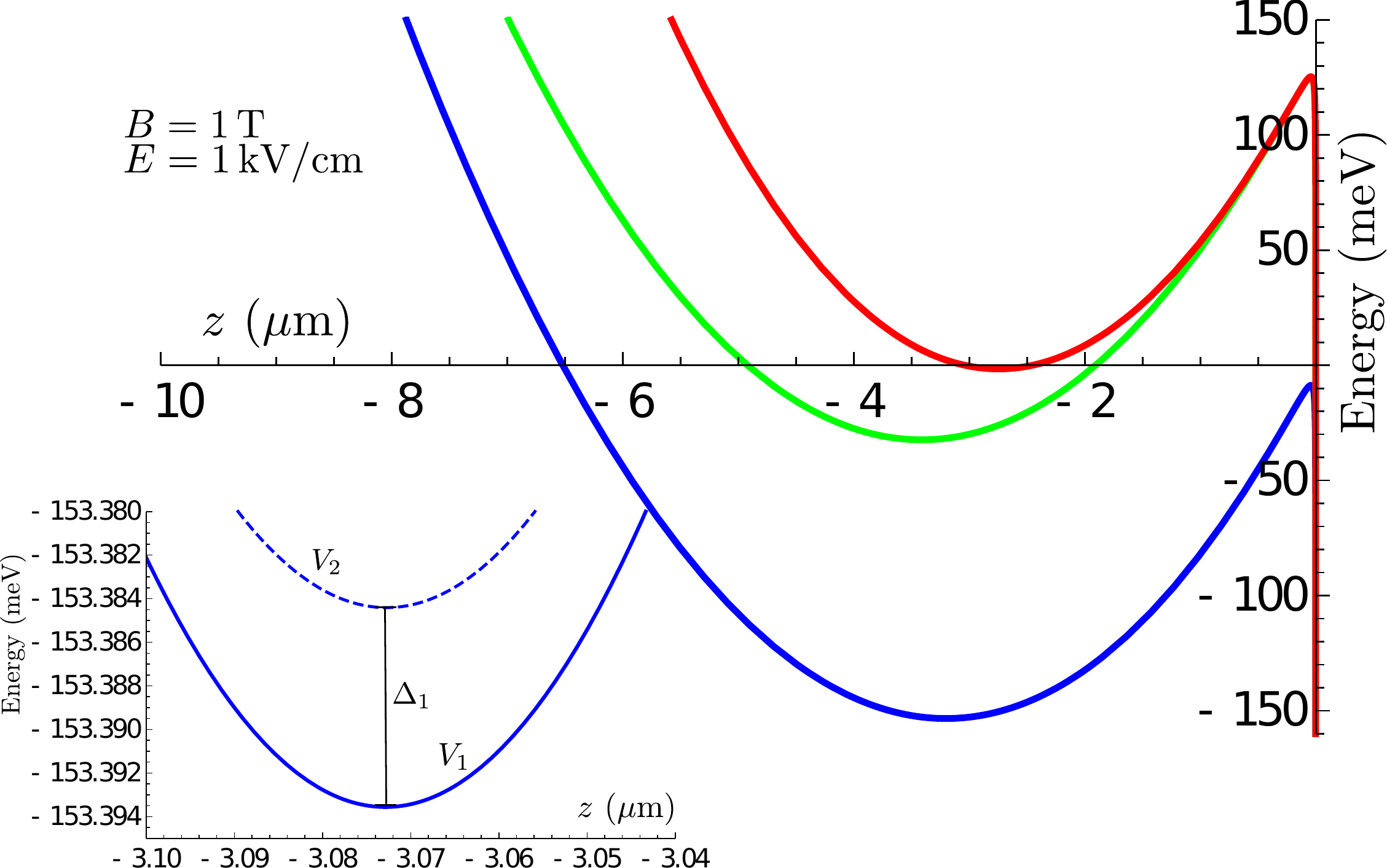} 
\caption{Potential curves for $B=1\, \rm T$, $E=1\, \rm kV/cm$. The specific 
field configurations are $\bfmath{B}||[1 0 0]$ and $\bfmath{E}||[001]$. The 
inset shows the adjacent potential curves $V_1$ and $V_2$ where the spacing 
$\Delta_1$ is indicated as well.}
\label{plot1a}
\end{figure}

As it has been shown in Ref.~\cite{Kurz2017}, the diagonalization of the 
giant-dipole potential $V_{\rm gd}$ provides six distinct potential energy 
surfaces with energetic separations in the range of a few hundred 
$\mu \textrm{eV}$ up to $100\, \textrm{meV}$. In Fig.~\ref{plot1a}, we show 
typical potential curves for field strengths $B=1\, \textrm{T}$ and $E=1\, 
\textrm{kV/cm}$. One clearly observes local potential minima at distances 
several micrometers away from the Coulomb center. For each potential 
surface, we obtain the corresponding eigenvector $| \phi_{i}(\bfmath{r}) 
\rangle,\ i=1,...,6$ including their spatial dependence on the electron-hole 
separation $\bfmath{r}$. 
We can define the following quantities that characterize the individual
giant-dipole potential curves:
\begin{itemize}
\item the potential depth $V^{(\alpha)}_{\rm d}$ given by
\begin{eqnarray*}
V^{(\alpha)}_{\rm d}=\lim_{x \rightarrow \infty} \ 
V_{\alpha}(x,0,z^{(\alpha)}_{\rm min})
-V_{\alpha}(\bfmath{r}^{(\alpha)}_{\rm min}),
\end{eqnarray*}
\item the quantity $\Delta_{\alpha}$ defining the energetic separation between 
two adjacent potential surface, i.e.
\begin{eqnarray*}
\Delta_{\alpha}=
V_{\alpha+1}(\bfmath{r}^{(\alpha+1)}_{\rm min})
-V_{\alpha}(\bfmath{r}^{(\alpha)}_ {\rm min}).
\end{eqnarray*}
\end{itemize}
We emphasize that all six potential surfaces possess local minima. This is 
in contrast to previous work \cite{Kurz2017}, where only four out of six 
surfaces possessed minima, which results from a different choice of Luttinger 
parameters which were only published recently \cite{Schweiner2016}. Indeed, the 
potential surfaces' topologies sensitively depend on the specific 
values of the Luttinger parameters \cite{Luttinger1954,Luttinger1956}. For this 
reason, a precise determination of excitonic giant-dipole properties such as 
level spacings and binding energies might provide the possibility of 
determining the specific Luttinger parameters with a higher degree of accuracy. 

Although the giant-dipole potential is diagonal within this basis, the set 
$\{|\phi_i(\bfmath{r}) \rangle_{i=1,...,6} \}$ is not suitable to diagonalize  
the total excitonic Hamiltonian $H_{\rm ex}$ as the kinetic part of 
Eq.~(\ref{Hex}) does not commute with the potential. More precisely, the 
coupling between different eigenstates $|\phi_i(\bfmath{r}) \rangle$ generated 
by the kinetic energy operator induces transitions between the potential energy 
surfaces $V_{i}(\bfmath{r})$. This feature is well-known in molecular physics 
where these kinds of non-adiabatic transitions between electronic eigenstates 
are induced by the kinetic energy of the nuclei 
\cite{Domcke_Conical_Intersections}. 

In a first ansatz, we follow the adiabatic approach from molecular physics by 
neglecting all excitonic transitions between a set of different potential 
surfaces. In particular, we define effective Hamiltonians 
\begin{eqnarray}
H^{(\alpha)}_{\rm eff} \equiv \langle \phi_{\alpha}(\bfmath{r}) |H_{\rm ex}| 
\phi_{\alpha}(\bfmath{r})\rangle_{\rm spin} \label{Hexeff},\ \alpha=1,...,6 
\label{H_eff_def}
\end{eqnarray}
by introducing
\begin{eqnarray}
I^{(\alpha)}_{i}(\bfmath{r}) &\equiv& \langle \phi_{\alpha}(\bfmath{r}) |I_{i}| 
\phi_{\alpha}(\bfmath{r}) \rangle_{\rm spin}, \ \ i=x,y,z,\\
\langle \Omega_{ki} \rangle_{\alpha}(\bfmath{r}) &\equiv& \langle 
\phi_{\alpha}(\bfmath{r}) |\Omega_{ki}| \phi_{\alpha}(\bfmath{r}) 
\rangle_{\rm spin},\\
\pi^{(\alpha)}_{i}(\bfmath{r})&\equiv& \langle \phi_{\alpha}(\bfmath{r}) 
|\pi_{i}| \phi_{\alpha}(\bfmath{r}) \rangle_{\rm spin}.
\label{spins_scalar_p}
\end{eqnarray}
In these expressions, the expectation values $\langle ... \rangle_{\rm spin}$ 
are only computed with respect to the spin-1 and spin-1/2 degrees of freedom 
$\bfmath{I}$ and $\bfmath{S}_h$, 
respectively. This means that the effective quantities 
$H^{(\alpha)}_{\rm eff}$ and $I^{(\alpha)}_{i}(\bfmath{r})$ are functions of 
the canonical conjugated variables $\bfmath{p}$ and $\bfmath{r}$, respectively. 
In particular, the components $\pi_i$ of the kinetic momentum are now given by
\begin{equation}
\pi^{(\alpha)}_{i}=p_i-qA^{(\alpha)}_{i}(\bfmath{r})
\end{equation}
with
\begin{equation}
A^{(\alpha)}_i=A_{i}(\bfmath{r})+g^{(\alpha)}_{i}(\bfmath{r})
\end{equation}
and
\begin{equation}
g^{(\alpha)}_{i}(\bfmath{r})=\frac{1}{q}\sum_{k}\left(\frac{m_h}{M}\delta_{ki}
-\langle \Omega_{ki} \rangle_{\alpha} \right) \tilde{K}_{k}. 
\end{equation}
Obviously, in the adiabatic approximation the homogeneous magnetic field is 
replaced by a spatially dependent field that can be computed from
\begin{equation}
\bfmath{B}^{(\alpha)}(\bfmath{r})= \bfmath{\nabla} \times 
\bfmath{A}^{(\alpha)}(\bfmath{r}). 
\end{equation}
By defining spatially dependent Luttinger parameters
\begin{eqnarray}
\gamma^{(\alpha)}_{2,i} \equiv \gamma_2 I^{(\alpha)2}_{i}(\bfmath{r}) ,\ \ 
\gamma^{(\alpha)}_{3,ij} \equiv \gamma_3 
I^{(\alpha)}_{i}(\bfmath{r})I^{(\alpha)}_{j}(\bfmath{r}), \label{Lutt_spatial}
\end{eqnarray}
we can write the effective Hamiltonians $H^{(\alpha)}_{\rm eff}$ as
\begin{eqnarray}
H^{(\alpha)}_{\rm eff}&=&
\frac{\bfmath{\pi}^2}{2\mu}-\frac{3}{\gamma^{\prime}_{1} } \left( 
\gamma^{(\alpha)}_{2,x}(\bfmath{r})\pi^{(\alpha)2}_{x}+\textrm{c.p.} \right) 
\nonumber\\ &\ &-\frac{6}{\gamma^{\prime}_{1}}
\left(\gamma^{(\alpha)}_{3,xy}(\bfmath{r})\pi^{
(\alpha)}_{x} \pi^{(\alpha)}_{y}+\textrm{c.p.} \right) 
+ V^{(\alpha)}_{\rm gd}(\bfmath{r})
\end{eqnarray}
with $\mu^{-1}=1+4\gamma_2/\gamma^{\prime}_{1}$ [see Eq.~(\ref{H_eff_def})].

\subsection{Strong electric-field limit}
\label{strong_e_field}
Before we analyze the excitonic system in adiabatic approximation, we consider 
the limit of strong electric fields. In this limit, one can neglect the 
spin-orbit coupling $H_{\rm so}$ as well as the magnetic field coupling 
$H_{B}$. In this approximation, the excitonic Hamiltonian reduces to the direct 
sum $H_{\rm ex}=H_{0} \oplus 1_{s=1/2}$. Hence, the problem of determining the 
excitonic giant-dipole states is equivalent to the eigenvalue problem of a 
$3 \times 3$-matrix, which can be solved analytically for arbitrary 
electric and magnetic field configurations. However, as it has been shown in 
Ref.~\cite{Kurz2017}, for a magnetic field oriented along the $[100]$ and an 
electric field in the $[001]$ direction, the expressions for the potential 
energy surfaces $V_{i}(\bfmath{r})$ and eigenstates $|\phi_{i}(\bfmath{r}) 
\rangle$ are more compact and given by
\begin{eqnarray}
V_1(\bfmath{r})&=&\left( \Omega_1- \Omega_2 \right)\tilde{K}^2 
+Ez-\frac{1}{r},\nonumber\\
V_{2,3}(\bfmath{r})&=&\left( \Omega_1- \Omega_2 \right)\tilde{K}^2 
+Ez-\frac{1}{r}+\frac{3}{2}\Omega_{2}\left( \tilde{K}^{2}_{2}+\tilde{K}^{2}_{3} 
\right)\nonumber \\ 
&\ & \pm \frac{1}{2}\sqrt{9\Omega^{2}_{2}\left( 
\tilde{K}^{2}_{2}-\tilde{K}^{2}_{3} \right)^2 + 
4\Omega^{2}_{3}\tilde{K}^{2}_{2}\tilde{K}^{2}_{3}}, 
\label{Vgd_analytic} 
\end{eqnarray}
and
\begin{eqnarray}
|\phi_1(\bfmath{r}) \rangle&=&|1\rangle,\nonumber \\ 
|\phi_2(\bfmath{r}) \rangle&=&\cos(\gamma)|2\rangle-\sin(\gamma)|3\rangle,\nonumber \\ 
|\phi_3(\bfmath{r}) \rangle&=&\sin(\gamma)|2\rangle+\cos(\gamma)|3\rangle 
\label{eigenstates_angle}
\end{eqnarray}
where the mixing angle $\gamma(\bfmath{r})$ is defined as
\begin{eqnarray}
\tan(2\gamma)=
\frac{2\Omega_3\tilde{K}_2\tilde{K}_3}{3\Omega_2(\tilde{K}^2_2-\tilde{K}^2_3)}.
\label{mixing_angle}
\end{eqnarray}
Interestingly, the mixing angle does not depend on the external electric field. 
In case that also $\bfmath{K}=0$, even the dependence on the magnetic field 
cancels out. If we calculate the quantities from Eq.~(\ref{spins_scalar_p}) in 
the strong electric field limit, we obtain
\begin{eqnarray*}
I^{(\alpha)}_{i}(\bfmath{r})=0,\ \ 
\langle \Omega_{ki} \rangle_{\alpha}&=&(C_1-\frac{2}{3}C_2)\delta_{ki}
\end{eqnarray*}
and
\begin{eqnarray}
\pi_{i}=p_i-q\tilde{A}^{(i)}_{\rm sym}(\bfmath{r})-\frac{\tilde{m}_{h}}{M}\tilde{K}_{i}, \label{pi_eff}
\end{eqnarray}
with
\begin{eqnarray}
\tilde{m}_h=m_h-M(C_1-\frac{2}{3}C_2)
\end{eqnarray}
and
\begin{eqnarray}
\tilde{A}^{(i)}_{\rm sym}(\bfmath{r})=\frac{1}{2}\tilde{\bfmath{B}} \times 
\bfmath{r},\ \ \ \tilde{\bfmath{B}}=\left(1+2\frac{\tilde{m}_h}{qM} \right) 
\bfmath{B}.\label{B_eff_strong_e}
\end{eqnarray}
The $\tilde{K}$-dependent term in Eq.~(\ref{pi_eff}) can be written as 
$\partial_i \tilde{K}_i x_i$, i.e. it can be eliminated by a simple gauge 
transformation. In addition, the giant-dipole Hamiltonian is determined by an effective 
magnetic field $\tilde{\bfmath{B}}$ that is parallel to the initial $B$-field, 
but possesses a different magnitude 
$\tilde{B}$ with $\tilde{B}/B=1+2\tilde{m}_h/qM \approx 1.6$, which is an 
enhancement of around $60 \%$. We note that both quantities $\tilde{m}_h$ and 
$\bfmath{\tilde{B}}$ do not depend on the specific potential surface. 

We finally obtain in the strong electric field approximation the following 
excitonic Hamiltonian 
\begin{eqnarray}
H^{(\alpha)}_{\rm eff}= \frac{\bfmath{\pi}^2}{2 \mu}+V_{\alpha}(\bfmath{r}), 
\label{Hexalpha}
\end{eqnarray}
whereby the potentials $V_{\alpha}$ are given by Eq.~(\ref{Vgd_analytic}). The 
set of effective excitonic Hamiltonians $H^{(\alpha)}_{\rm eff}$ is 
identical to the Hamiltonian discussed previously \cite{Kurz2017}. There it has been shown that the giant-dipole potential surfaces 
$V_{\alpha}(\bfmath{r})$ possess minima at $\bfmath{r}_{\rm min}=(0,0,z_{\rm min})$ with 
\begin{eqnarray*}
z^{(1,2)}_{\rm min}&=&\frac{E}{6(\Omega_1+\frac{1}{2}(1 \pm 3)\Omega_2)B^2}[2\cos(\frac{\theta+2\pi}{3})-1]
\end{eqnarray*}
and
$\cos(\theta)=54(\Omega_1+\frac{1}{2}(1 \pm 3)\Omega_2)^2B^4/E^3-1$. Although 
Eq.~(\ref{Hexalpha}) is very similar to the atomic Hamiltonian discussed in 
Ref.~\cite{Dippel1994}, we stress that in the present case the 
potential $V_{\alpha}(\bfmath{r})$ is determined by the bare external magnetic 
field $\bfmath{B}$, while the kinetic energy term in Eq.~(\ref{Hexalpha}) 
depends on the effective field $\tilde{\bfmath{B}}$.

In order to obtain the energies and wave functions of the excitonic 
giant-dipole species, we expand the potential surfaces around their local 
minima. Including terms up to second order, we find the harmonically 
approximated potentials
\begin{eqnarray}
V^{(\alpha)}_{\rm h}(\bfmath{r})=\frac{\mu}{2}\omega^{(\alpha)}_{x}x^2 + 
\frac{\mu}{2}\omega^{(\alpha)}_{y}y^2 + \frac{\mu}{2}\omega^{(\alpha)}_{z}z^2 
\end{eqnarray}
with the frequencies
\begin{eqnarray}
\omega^{(\alpha)}_{x}&=&\sqrt{-\frac{1}{\mu z^{(\alpha)3}_{\rm min}}},\ \ 
\alpha=1,2,3\ \ , \nonumber\\
\omega^{(1)}_{y}&=&\omega^{(3)}_{y}=\sqrt{\frac{1}{\mu}
\left(2(\Omega_1-\Omega_2)B^2-\frac{1}{ z^{(1)3}_{\rm min}}\right)}, \nonumber\\
\omega^{(2)}_{y}&=&\sqrt{\frac{1}{\mu}\left(2(\Omega_1+2\Omega_2)B^2
-\frac{1}{ z^{(2)3}_{\rm min}}\right)}, \nonumber\\
\omega^{(1)}_{z}&=&\omega^{(3)}_{z}=\sqrt{\frac{2}{\mu}
\left((\Omega_1- \Omega_2)B^2+\frac{1}{ z^{(1)3}_{\rm min}}\right)}, \nonumber\\
\omega^{(2)}_{z}&=&\sqrt{\frac{2}{\mu}\left((\Omega_1+2\Omega_2)B^2
+\frac{1}{ z^{(2)3}_{\rm min}}\right)}.
\end{eqnarray}
As it has been discussed in Ref.~\cite{Kurz2017}, the eigenenergies 
and eigenstates can be obtained analytically via a unitary transformation which 
decouples the $(y,z)$-degrees of freedom leading to a set of three decoupled 
harmonic oscillators. Apart from the frequencies $\omega^{(\alpha)}_{z}$, the 
remaining energy spacings are equidistant with frequencies   
\begin{eqnarray*}
\omega^{(\alpha)}_{1,2}&=&\frac{1}{\sqrt{2}}
[\omega^{(\alpha)2}_{z}+\omega^{(\alpha)2}_{y}+\omega^{2}_{c}\\ &\ &\pm 
\sqrt{(\omega^{(\alpha)2}_{z}+\omega^{(\alpha)2}_{y}
+\omega^{2}_{c})^2-4\omega^{(\alpha)2}_{z}\omega^{(\alpha)2}_{y}}]^{1/2}
\end{eqnarray*}
and $\omega_c=q \tilde{B}/\mu$.

\begin{figure}[ht]
\centering
\begin{minipage}{0.425\textwidth} 
\includegraphics[width=\textwidth]{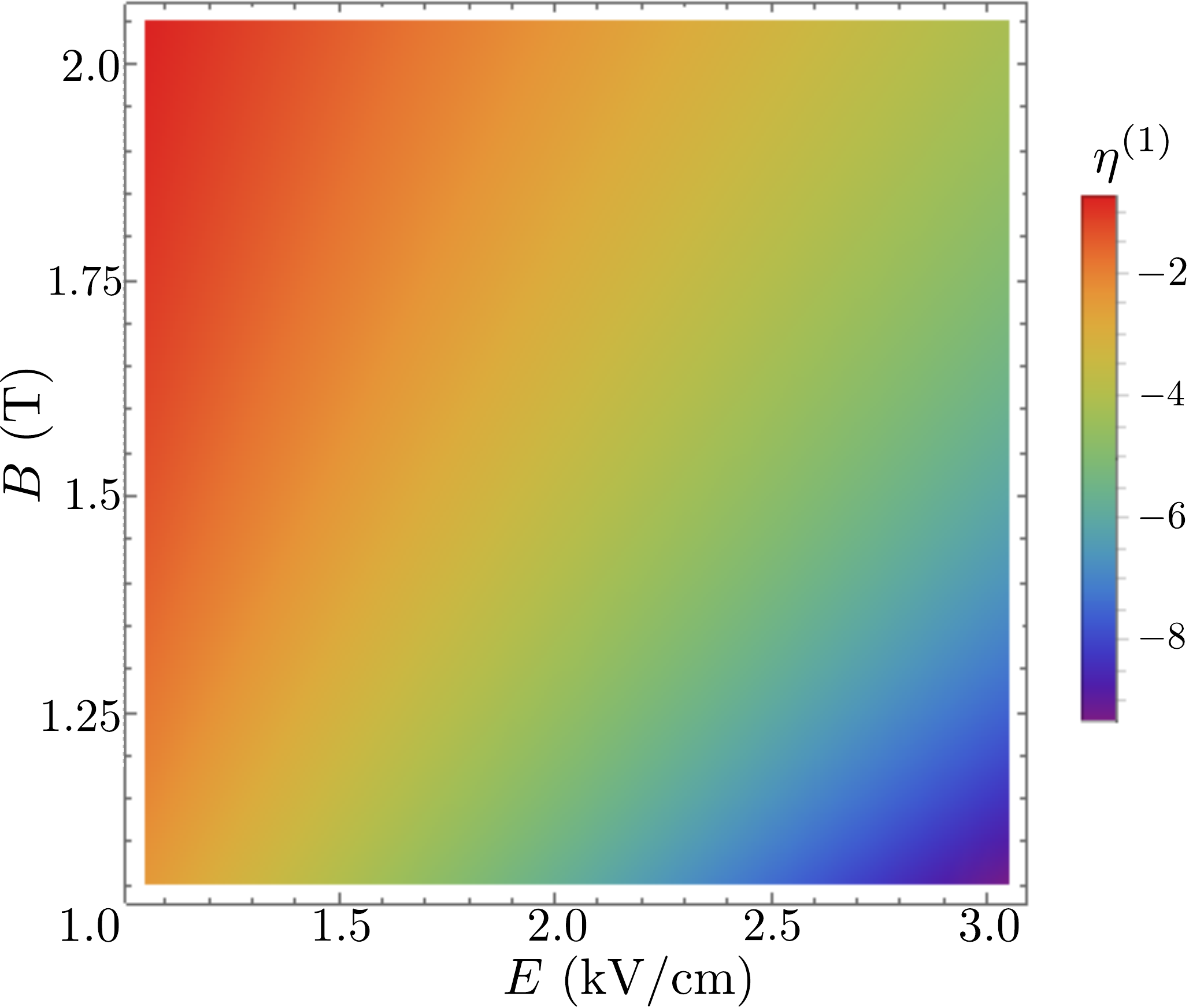}  
\end{minipage}
\caption{Density plot of $\eta^{(1)}$ for 
$1\, \textrm{kV/cm} \le E \le 3\, \rm kV/cm$ and 
$1\, \rm{T} \le B \le 2\, \rm T$. We see that in this strong-field limit we 
have $\eta^{(1)}<0$, i.e. no bound states are found in this regime.}
\label{eta1}
\end{figure}

Although the excitonic eigenenergies and eigenstates are given analytically, one
has to remember that these results have been derived within an harmonic 
approximation in the vicinity of the outer potential well. However, the exact 
potential surfaces possess an ionization limit in the direction of the external 
magnetic field. For this reason, one has to ensure that for a certain field 
configuration the calculated ground state still lies deep within the outer 
potential well. To analyze this issue in more detail we define the quantity
\begin{eqnarray}
\eta^{(\alpha)}&\equiv&
\frac{V^{(\alpha)}_{\rm d}-(\varepsilon_{000}-V_{\rm min})}
{V^{(\alpha)}_{\rm d}}\nonumber \\
&=&1-\frac{|z^{(\alpha)}_{\rm min}|
(\omega^{(\alpha)}_{x}+\omega^{(\alpha)}_{1}+\omega^{(\alpha)}_{2})}{2}, 
\end{eqnarray}
which accounts for the energy difference between the potential depth 
$V^{(\alpha)}_{\rm d}$ and the spacing between the ground state and the 
potential minimum, i.e. $\varepsilon_{000}-V_{\rm min}$. 
In Fig.~\ref{eta1}, we show the quantity $\eta^{(1)}$ for electric and 
magnetic fields of $1\, \textrm{kV/cm} \le E \le 3\, \textrm{kV/cm}$ 
and $1\, \textrm{T} \le B \le 2\, \textrm{T}$, respectively. 
One observes that in the considered field strength regime $\eta^{(1)}<0$, which 
means that the giant-dipole ground state lies above the ionization limit of the 
potential surface. The same result are obtained for the remaining potential 
surfaces, i.e. $\eta^{(2,3)}<0$. As a consequence, we expect no bound 
excitonic giant-dipole states in the limit of strong electric fields.

\subsection{Arbitrary field strengths \label{arbitrary_fields}}
In a next step, we keep the adiabatic approximation but leave the limit of 
strong electric fields in order to analyze arbitrary field strengths. Again, we 
consider $\bfmath{B}||[100]$ and $\bfmath{E}||[001]$. In this case, a rigorous 
analysis is rather complicated as the adiabatic Hamiltonians 
$H^{(\alpha)}_{\rm eff}$ do not only depend on spatially varying magnetic 
fields, but also on spatially dependent Luttinger parameters defined in 
Eq.~(\ref{Lutt_spatial}). However, we may employ the fact that we are mainly 
interested in the bound states localized around the minima of the outer 
potential wells. For this reason, we make use of the approximation that the 
eigenstates do not strongly vary in the vicinity of a certain potential minimum. 
To illustrate that in more detail, we go back to the strong-field limit 
discussed in the previous section. According to Eq.~(\ref{eigenstates_angle}),
the spatial dependence of the eigenvectors are given by the mixing angle 
$\gamma$ determined by Eq.~(\ref{mixing_angle}). If we consider $K=0$, we 
directly see that $\gamma(\bfmath{r}_{\rm min})=0$, which gives
\begin{eqnarray}
|\phi_1(\bfmath{r}_{\rm min}) \rangle &=&|1\rangle,\nonumber\\ 
|\phi_2(\bfmath{r}_{\rm min}) \rangle&=&|2\rangle,\nonumber\\  
|\phi_3(\bfmath{r}_{\rm min}) \rangle&=&|3\rangle.
\end{eqnarray}
Obviously, the eigenstate $|\phi_1(\bfmath{r}) \rangle=
|\phi_1(\bfmath{r}_{\rm min})=| 1 \rangle\ \forall\ \bfmath{r} \in 
\mathbb{R}^{3}$. To analyze the deviation of the remaining eigenstates from the 
corresponding eigenstates at the minimum positions, we need to look at the 
spatial dependence of $\cos(\gamma(\bfmath{r}))$ in more detail.
\begin{figure}[ht]
\centering
\begin{minipage}{0.455\textwidth} 
\includegraphics[width=\textwidth]{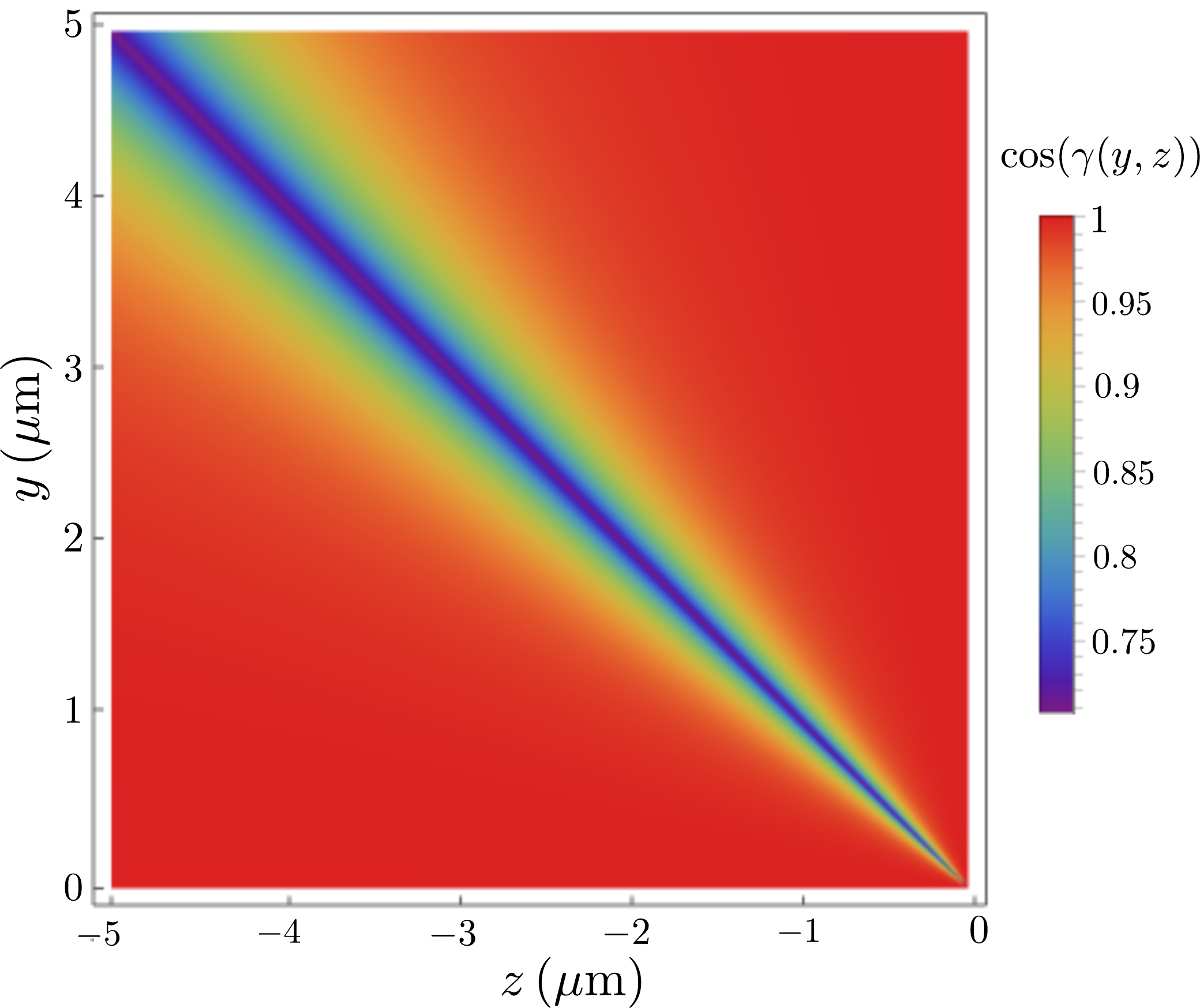}  
\end{minipage}
\caption{Density plot of $\cos(\gamma(y,z))$ for applied field strengths of 
$E=1\, \textrm{kV/cm}$ and $B=1\, \textrm{T}$. In the vicinity of the potential 
minima $z \approx -3\, \mu\textrm{m}$, one finds 
$\cos(\gamma(y,z))\approx 1$, which means that the corresponding eigenvectors 
only minor differ from the eigenstate at the potential minimum.}
\label{plot_cos}
\end{figure}

In Fig.~\ref{plot_cos}, we show the mixing angle $\cos(\gamma(y,z))$ for 
applied field strengths $B=1\, \textrm{T}$ and $E=1\, \textrm{kV/cm}$ in the 
spatial range $0 \le y \le 5\, \mu \rm m$ and $-5\, \mu \textrm{m} \le z \le 0$. 
For these field strengths, the potential minima are located at 
$z^{(1)}_{\rm min}=-3.07\, \mu \textrm{m}$,  $z^{(3)}_{\rm min}=-3.25\, \mu 
\textrm{m}$ and $z^{(5)}_{\rm min}=-2.8\, \mu \textrm{m}$, respectively. 
We see that, in the vicinity of the potential minima $(z \approx 3\, \mu 
\textrm{m})$, $\cos(\gamma)$ remains close to unity, which means 
that the deviations from the pure eigenstates close to the minimum positions 
remain negligible. This result is not only valid for strong electric fields, but also for all 
field strengths considered throughout this work. This means that for the 
calculation of the matrix elements of the kinetic energy we can use the 
eigenvectors at the minimum position, i.e. $I^{(\alpha)}_{i} 
(\bfmath{r})\rightarrow I^{(\alpha)}_{i}(\bfmath{r}_{\rm min})$. 

Analogous to the strong electric-field limit, we now define a renormalized 
vector potential using Eq.~(\ref{spins_scalar_p}), from which the effective 
magnetic field $\bfmath{\tilde{B}}^{(\alpha)}$ is obtained as
$\bfmath{\tilde{B}}^{(\alpha)}=\grad \times \bfmath{\tilde{A}}^{(\alpha)}$. 
However, it turns out that
\begin{equation}
\sum_{k}\langle \Omega_{ki} \rangle_{\alpha} \approx \langle \Omega_{ii} \rangle_{\alpha}, 
\end{equation}
which means that the components $\tilde{B}^{(\alpha)}_i$ of the effective 
magnetic field are given by
\begin{equation}
\tilde{B}^{(\alpha)}_{i}=\left[ 1+\frac{2}{q} \left( \frac{m_h}{M}-\langle 
\Omega_{ii} \rangle_{\alpha}\right) \right]B_i.
\end{equation}
The magnetic field $\bfmath{\tilde{B}}^{(\alpha)}$ is, in general, no 
longer parallel to the incident field $\bfmath{B}$ as it now points into the 
direction of the unit vector 
\begin{equation}
\bfmath{e}^{(\alpha)}_{\tilde{\bfmath{B}}} =
\frac{1}{|\tilde{\bfmath{B}}^{(\alpha)}|}
(\tilde{B}^{(\alpha)}_{1},\tilde{B}^{(\alpha)}_{2},\tilde{B}^{(\alpha)}_{3})^T,
\end{equation}
with the magnitude $|\tilde{\bfmath{B}}^{(\alpha)}|=
\sqrt{\tilde{B}^{(\alpha)2}_1+\tilde{B}^{(\alpha)2}_2+\tilde{B}^{(\alpha)2}_3}$. 
In contrast to the strong electric-field limit discussed in 
Sec.~\ref{strong_e_field}, the effective magnetic field now depends on the 
specific potential curve under consideration via the matrix elements of the 
$\Omega_{ii}$-matrices. 

\begin{figure}[ht]
\centering
\begin{minipage}{0.45\textwidth} 
\includegraphics[width=\textwidth]{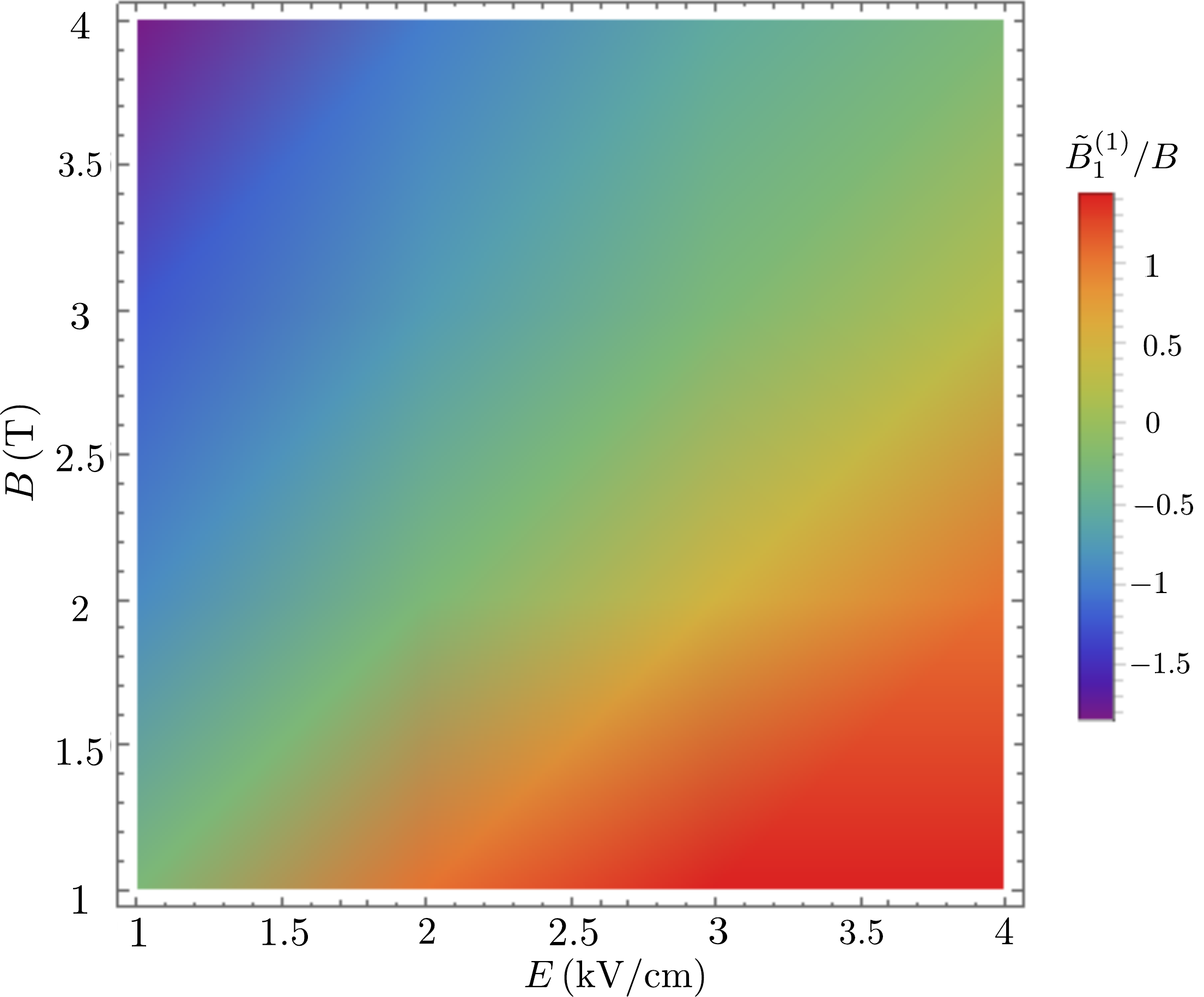} 
\end{minipage}
\caption{Effective magnetic field component $\tilde{B}^{(1)}_{1}/B$ for 
external field strengths $1\, \textrm{kV/cm} \le E \le 4\, \textrm{kV/cm}$,
$1\, \textrm{T}\le B \le 4\, \textrm{T}$.}
\label{Beff}
\end{figure}

In Fig.~\ref{Beff}, we show $\tilde{B}^{(1)}_1/B$ as a function of the 
external field parameters for applied field strengths in the range of 
$1\,\textrm{T}\le B\le 4\,\textrm{T}$ and 
$1\,\textrm{kV/cm}\le E\le 4\,\textrm{kV/cm}$. In contrast to the approximation 
discussed in Sec.~\ref{strong_e_field} we now find that, depending on the 
specific field strengths, not only positive-valued effective magnetic field 
strengths, but also fields with negative values. In particular, for strong 
external magnetic fields one finds $\tilde{B}^{(1)}_{1} <0$, which means that 
not only the magnitude of the magnetic field is modified but also its direction 
with respect to the external field is changed. However, for sufficiently strong 
electric fields the sign of the magnetic field becomes positive and reaches a 
maximum value of around $1.5B$. As expected, this is quite close to the 
effective $B$-field $\tilde{B}=1.6B$ obtained in the strong electric-field approximation derived in 
Eq.~(\ref{B_eff_strong_e}).

In order to calculate the excitonic spectrum within this approximation, we use 
the renormalized Luttinger parameters to define effective masses 
$\mu^{(\alpha)}_{i},\ i=x,y,z$, as
\begin{eqnarray}
\frac{1}{\mu^{(\alpha)}_{i}}&\equiv& 
1+4\frac{\gamma_2}{\gamma^{'}_{1}}\left[1-\frac{3}{2}
\gamma^{(\alpha)}_{2,i}(\bfmath{r}_{\rm min})\right], \nonumber\\
\gamma^{(\alpha)}_{ij}&\equiv&
-\frac{6\gamma_3}{\gamma^{'}_{1}}
\gamma^{(\alpha)}_{3,ij}(\bfmath{r}_{\rm min}). 
\end{eqnarray}
This means that for all three potential curves we obtain effective 
Hamiltonians $H^{(\alpha)}_{\rm eff}$ with
\begin{eqnarray}
H^{(\alpha)}_{\rm eff}&=&
\sum_{i}\frac{\pi^{(\alpha)2}_{i}}{2\mu^{(\alpha)}_{i}}
+\sum_{i \not= j}\gamma^{(\alpha)}_{ij}\pi^{(\alpha)}_{i}\pi^{(\alpha)}_{j} 
+V_{\alpha}(\bfmath{r}), \label{htr} \nonumber\\
\pi^{(\alpha)}_{i}&=&p_i-q\tilde{A}^{(\alpha)}_{i}.
\end{eqnarray}
Analogous to the strong electric-field approximation, the exact interaction 
potentials $V_{\alpha}(\bfmath{r})$ can be expanded around their minimum
positions $\bfmath{r}^{(\alpha)}_{\rm min}$. By defining the frequencies 
\begin{equation}
C^{(\alpha)}_{ij}\equiv\frac{\partial^2}{\partial x_{i} \partial x_j}
V_{\alpha}(\bfmath{r})|_{\bfmath{r}=\bfmath{r}_{\rm min}},
\end{equation}
the exact potentials can be approximated by
\begin{eqnarray*}
V^{(\alpha)}_{\rm h}(\bfmath{r})&=&V^{(\alpha)}_{\rm min}+\frac{1}{2}
\bigg(C^{(\alpha)}_{xx}x^2 + C^{(\alpha)}_{yy}y^2 +C^{(\alpha)}_{zz}z^2  \\
&+&C^{(\alpha)}_{xy}xy+C^{(\alpha)}_{xz}xz+C^{(\alpha)}_{yz}yz \bigg)
\end{eqnarray*}
with $V^{(\alpha)}_{\rm min} \equiv V_{\alpha}(\bfmath{r}^{(\alpha)}_{\min})$. 
Together with the $\pi^{(\alpha)}_i$-dependent terms in Eq.~(\ref{htr}), the 
effective excitonic Hamiltonian is bilinear in the spatial and canonical 
momentum coordinates $\bfmath{r}$ and $\bfmath{p}$, respectively, and can thus
be written as
\begin{equation}
H^{(\alpha)}_{\rm eff}=\bfmath{x}^{T} \mathcal{H}^{(\alpha)}_{\rm eff} \bfmath{x},\ \ 
\bfmath{x}=(x\ y\ z\ p_x \ p_y\ p_z)^{T}\label{H_eff_w}
\end{equation}
in which $\mathcal{H}^{(\alpha)}$ is a $(6 \times 6)$-dimensional real, 
symmetric and positive definite matrix (see App.~\ref{app_williamson}). We 
note that, if the terms $\gamma^{(\alpha)}_{ij},C^{(\alpha)}_{ij},\ i \not= j$ 
are negligible, the effective Hamiltonians $H^{(\alpha)}_{\rm eff}$ are sums 
of three harmonic oscillators of charge $q$ and masses $\mu^{(\alpha)}_{i}$ in 
external effective magnetic fields. This problem can be solved by applying an 
unitary transformation that decouples the different degrees of freedom, and 
where the spectrum is determined by the coefficients $C^{(\alpha)}_{ii}$ 
\cite{Schmelcher1993a,Dippel1994}. 

In order to calculate the eigenenergies of Eq.~(\ref{H_eff_w}) exactly, we 
apply Williamson's theorem \cite{Wiliamson1936,Ikramov2018} which states that 
there exists a symplectic matrix $S^{(\alpha)} \in \textrm{Sp}(6,\mathbb{R})$ such that 
\begin{equation}
\mathcal{H}^{(\alpha)}_{\rm eff}=S^{(\alpha)} D^{(\alpha)} S^{(\alpha)T}\label{Heff_will}
\end{equation}
with 
$D^{(\alpha)}=\textrm{diag}(\lambda^{(\alpha)}_{1},\lambda^{(\alpha)}_{2},\lambda^{(\alpha)}_{3},
\lambda^{(\alpha)}_{1}, \lambda^{(\alpha)}_{2},\lambda^{(\alpha)}_{3})$, $\lambda^{(\alpha)}_{i=1,2,3}>0$. 
Importantly, the components of the transformed coordinate vector 
$\bfmath{q}^{(\alpha)}=S^{(\alpha)} \bfmath{x}$ fulfill the same commutation
relation as the $\bfmath{x}_i$, i.e.
\begin{eqnarray}
[q^{(\alpha)}_{i},q^{(\alpha)}_{j}]=iJ_{ij},\ \ 
J=\begin{bmatrix}
0 & 1_n\\
-1_n &0
\end{bmatrix}
\end{eqnarray}
where $1_n$ denotes the $n \times n$ unit matrix. In the new coordinates, the 
Hamiltonian $H^{(\alpha)}_{\rm eff}$ is given by
\begin{equation}
H^{(\alpha)}_{\rm eff}=\sum^{3}_{i=1}\lambda^{(\alpha)}_i 
\left[q^{(\alpha)2}_{i}+q^{(\alpha)2}_{i+2}\right], 
\label{Heff_trans} 
\end{equation}
i.e., we find three uncoupled harmonic oscillators with frequencies
$\tilde{\omega}^{(\alpha)}_i=2 \lambda^{(\alpha)}_{i},\ i=1,2,3$. This means 
that the eigenenergies of $H^{(\alpha)}_{\rm eff}$ are
\begin{equation}
\varepsilon^{(\alpha)}_{n_1n_2n_3}=\sum^{3}_{i=1}\tilde{\omega}^{(\alpha)}_{i}
(n_{i}+\frac{1}{2}),\ \tilde{\omega}^{(\alpha)}_{i} \equiv 
2\lambda^{(\alpha)}_i, \ \ n_{i}\in \mathbb{N}_0. 
\end{equation}
Similar to a three-dimensional harmonic oscillator, the energy 
spectrum is determined by three separate energy spacings 
$\tilde{\omega}^{(\alpha)}_{i},\ i=1,2,3$.

\begin{figure}[ht]
\centering
\begin{minipage}{0.425\textwidth} 
\includegraphics[width=\textwidth]{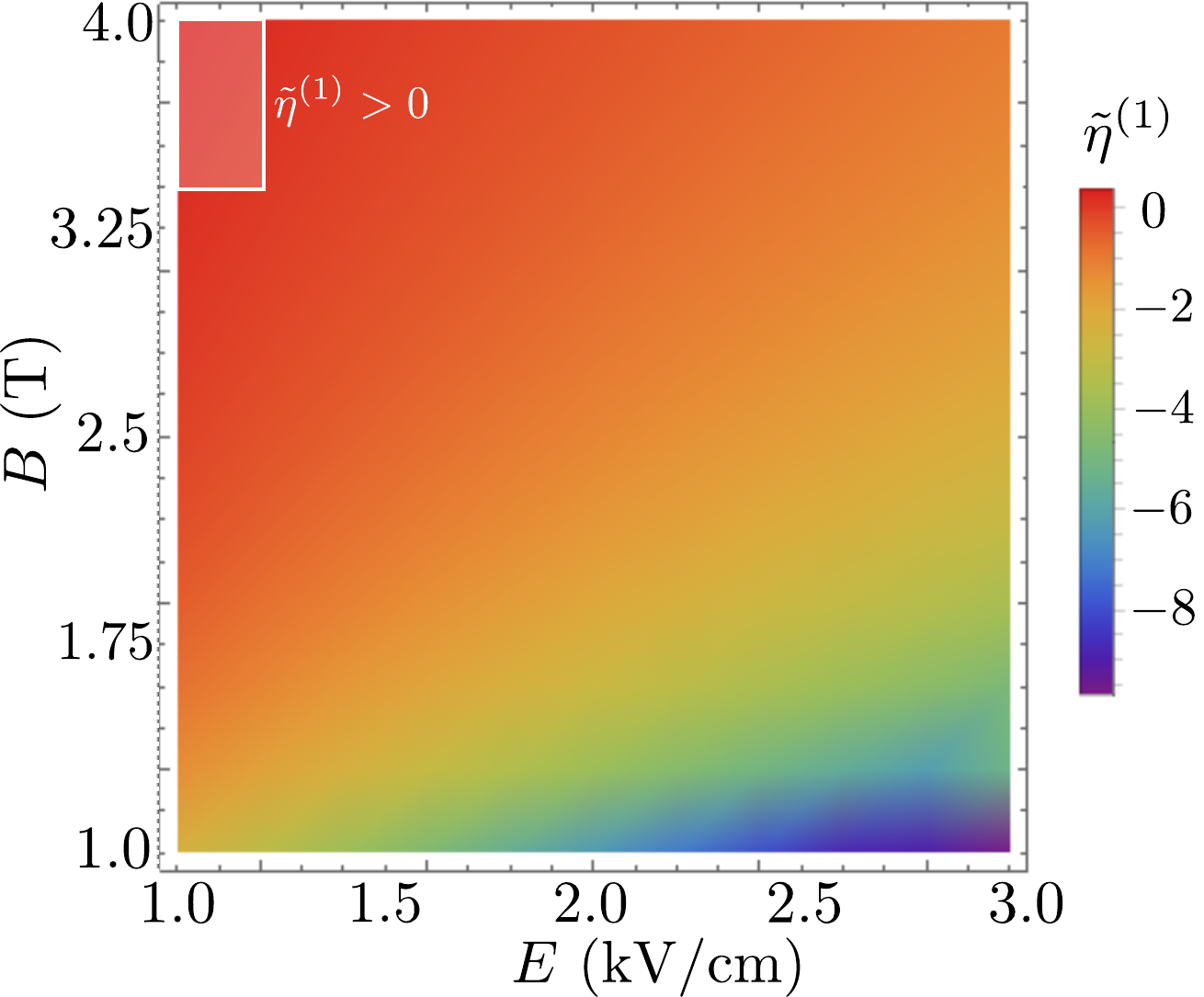}  
\end{minipage}
\caption{Density plot of $\tilde{\eta}^{(1)}$ for $1\, \rm kV/cm \le E \le 3\, 
\rm kV/cm$ and $1\, \rm{T} \le B \le 4\, \rm T$. For large electric fields,
$\tilde{\eta}^{(1)}<0$, while for strong magnetic fields one finds 
$\tilde{\eta}^{(1)}>0$. This region is indicated by the shaded box.}
\label{ionlimit}
\end{figure}

Analogous to the strong electric-field limit, we define the quantity 
$\tilde{\eta}^{(\alpha)}$ that measures the energy difference between the 
potential depth and the ground-state energy, and obtain
\begin{equation}
\tilde{\eta}^{(\alpha)}=1-\frac{1}{2V^{(\alpha)}_{\rm d}}\sum\limits_{i=1}^3
\tilde{\omega}^{(\alpha)}_i.
\end{equation}
In Fig.~\ref{ionlimit}, we show the energy difference $\tilde{\eta}^{(1)}$ for 
applied field strengths of $1\, \textrm{kV/cm} \le E \le 4\, \textrm{kV/cm}$ and 
$1\, \textrm{T} \le B \le 4\, \textrm{T}$. As before, one observes that, for increasing electric field strength,
$\tilde{\eta}^{(1)}$ becomes negative, which means that the ground state energy 
lies above the potential depth, i.e. no bound state is present. However, for 
low electric fields and high magnetic fields of around $B =4\, \textrm{T}$, we 
find a regime in which $\tilde{\eta}^{(1)}$ is positive. In Fig.~\ref{ionlimit},
this region is indicated by the shaded box. This result is reasonable as 
for increasing magnetic field strengths the potential depth increases as well. 
This means that, for sufficiently strong magnetic fields, the potential wells 
are deep enough to support bound excitonic giant-dipole states.

In Tab.~\ref{table2}, we present the frequencies 
$\tilde{\omega}^{(\alpha)}_{k}$ for the first, third and fifth potential 
surface for field strengths of $B=4\, \textrm{T}$ and $E=1\, \textrm{kV/cm}$. 
The frequencies for $\alpha=2,4,6$ are not explicitly shown as they are close 
to the values for the adjacent potential surface. One observes that, for all 
potentials one obtains two frequencies with values in the range of 
$87\,\mu\textrm{eV}-242\, \mu\textrm{eV}$. The third frequency is always larger 
and lies between $949\,\mu \textrm{eV}$ for the third and $1.58\,\textrm{meV}$ 
for the first potential curve. Compared to the potential depths of the 
corresponding surfaces, e.g.\ $V^{(1)}_{\rm d}=1.05\, \textrm{meV}$, the 
frequencies with $k=3$ are rather large, meaning that it is only possible to 
excite one state in the corresponding mode before one exceeds the potential 
depth and the harmonic approximation breaks down. However, as the frequencies 
of the remaining modes are smaller, one can easily excite a few states that
are still bound deep within the potential surfaces. 
\begin{table}[ht]
\centering
\begin{tabular}{ |c|c c c| }
\hline
\diagbox{$\alpha$}{$k$}
& $1$ & $2$ & $3$\\
\hline
$1$&$87$ & $142$ & $1580$ \\
$3$&$102$ & $161$ & $1366$\\
$5$&$114$ & $242$ & $949$ \\
\hline  
\end{tabular}
\caption{Excitonic frequencies $\tilde{\omega}^{(\alpha)}_{k}$ in 
$\mu\textrm{eV}$ for $B=4\, \textrm{T},\ E=1\, \textrm{kV/cm}$.}
\label{table2}
\end{table}
If we compare the frequencies $\tilde{\omega}^{(\alpha)}_{k}$ with the 
energetic spacing $\Delta_{\alpha}$, we find that 
$\textrm{min}(\tilde{\omega}^{(\alpha)}_{k=1,2,3}) \gg \Delta_{\alpha}\ \forall 
\alpha$, i.e. the frequencies are much larger than the spacing between adjacent 
potential surfaces. For this reason, we exspect a strong mixing between the 
potential surfaces in the case we include intrasurface couplings.  

Finally, we can determine the eigenstates by using the fact that the 
transformed effective Hamiltonian, Eq.~ (\ref{Heff_trans}), is a sum of three 
decoupled harmonic oscillators. Thus, we can construct ladder operators 
$a^{(\alpha)(\dagger)}_{k},\ k=1,2,3$ as
\begin{equation}
a^{(\alpha)(\dagger)}_{k}=\frac{1}{2}\left( q^{(\alpha)}_{k} \pm i 
q^{(\alpha)}_{k+2} \right).
\end{equation}
From here, the giant-dipole eigenstates $|n_1n_2n_3 \rangle$ are constructed via
\begin{gather}
| n_1 n_2 n_3 \rangle_{\alpha} =\frac{1}{\sqrt{n_1! n_2 ! n_3!}} \nonumber\\ 
\times \left( a^{(\alpha)\dagger}_{1}\right)^{n_1}
\left( a^{(\alpha)\dagger}_{2}\right)^{n_2}
\left( a^{(\alpha)\dagger}_{3}\right)^{n_3}| 0 \rangle.
\end{gather}
Using the transformation matrix $S^{(\alpha)}$ from Eq.~(\ref{Heff_will}), both the spatial and momentum coordinates can be expressed in terms of the ladder operators and vice versa. 
\section{Full excitonic spectra}
\label{full_spectra}
In order to provide a full analysis of the excitonic spectra, we next consider 
a full diagonalization approach to calculate the corresponding excitonic 
eigenenergies and states. Here, we are mostly interested in the determination of 
the ground state and the lowest lying giant-dipole states. In order to compute 
them most efficiently, one has to choose a basis set adapted to the properties 
of the system. As we are interested in the properties of the 
potentially bound excitonic giant-dipole states that are localized in the outer 
potential wells, it is clear that one should choose a set of basis functions 
that are also localized around the outer potential minima of the giant-dipole 
potential surfaces. As we have seen in Sec.~\ref{adiabatic_approx} in case of 
the adiabatic approximation, one obtains a set of effective giant-dipole Hamiltonians 
that can be diagonalized separately from one another. In this case, one obtains 
a set of basis functions that inherently possess the desired properties 
required for them to be a good choice for the diagonalization procedure. 
However, as we are interested in the lowest lying giant-dipole states, we use 
the fact that for the considered field strength regime the lowest potential 
energy surface is mostly determined by the first term in the expression of the 
excitonic giant-dipole potential [see Eq.~(\ref{Vgd})]. Expanding this term up 
to second order around the minimum position $z_{\rm{min}}$, i.e.\
\begin{gather}
\Omega_1 B^2(y^2+z^2)+Ez-\frac{1}{r} \nonumber\\
\approx V_{\rm{min}}+ \frac{1}{2}\omega^{2}_{x}x^2 
+\frac{1}{2}\omega^{2}_{y}y^2+\frac{1}{2}\omega^{2}_{z}(z-z_{\rm{min}})^2 
\label{V_approx},
\end{gather}
we define the basis functions $|\psi_{n_x,n_1,n_2} \rangle$ for our 
diagonalization procedure to be the giant-dipole eigenfunctions of a single 
particle of mass $\mu=1$, charge $q$, trapping frequencies $\omega_{i=x,y,z}$ 
and external fields $\bfmath{B}$ and $\bfmath{E}$, respectively. Together with 
the basis states of the spin-1 and spin-1/2 Hilbert space $|1,m \rangle \otimes 
|1/2,m_s \rangle,$ we define the following basis states
\begin{gather}
|\gamma,m,m_s \rangle \equiv |\psi_{n_x,n_1,n_2} \rangle \otimes |1,m \rangle 
\otimes |\frac{1}{2},m_s \rangle \\
n_{i}=0,1,2,... \ \ \ m=0, \pm 1 \ \ \ m_s= \pm \frac{1}{2}.
\end{gather}

For the exact diagonalization scheme we have calculated the matrix elements of 
the excitonic Hamiltonian (\ref{Hex_irr}), where the giant-dipole 
potential is approximated by Eq.~(\ref{V_approx}). 
Together with the six-dimensional spin-space, one obtains a 
$6(N_{x,\rm{max}}+1)(N_{1,\rm{max}}+1)(N_{2,\rm{max}}+1)$-dimensional matrix 
representation for the excitonic giant-dipole Hamiltonian, where 
$N_{i,\rm{max}}$ denote the maximal number of basis functions used within the
chosen basis set. Throughout our analysis, we obtained sufficient numerical 
convergence using $N_{x,\rm{max}}=N_{1,\rm{max}}=15,\ N_{2,\rm{max}}=5$, which 
yields a basis set of $9216$ states. 
\subsection{Magnetic field in [100]-direction}
\begin{figure}[ht]
\centering
\includegraphics[width=0.475\textwidth]{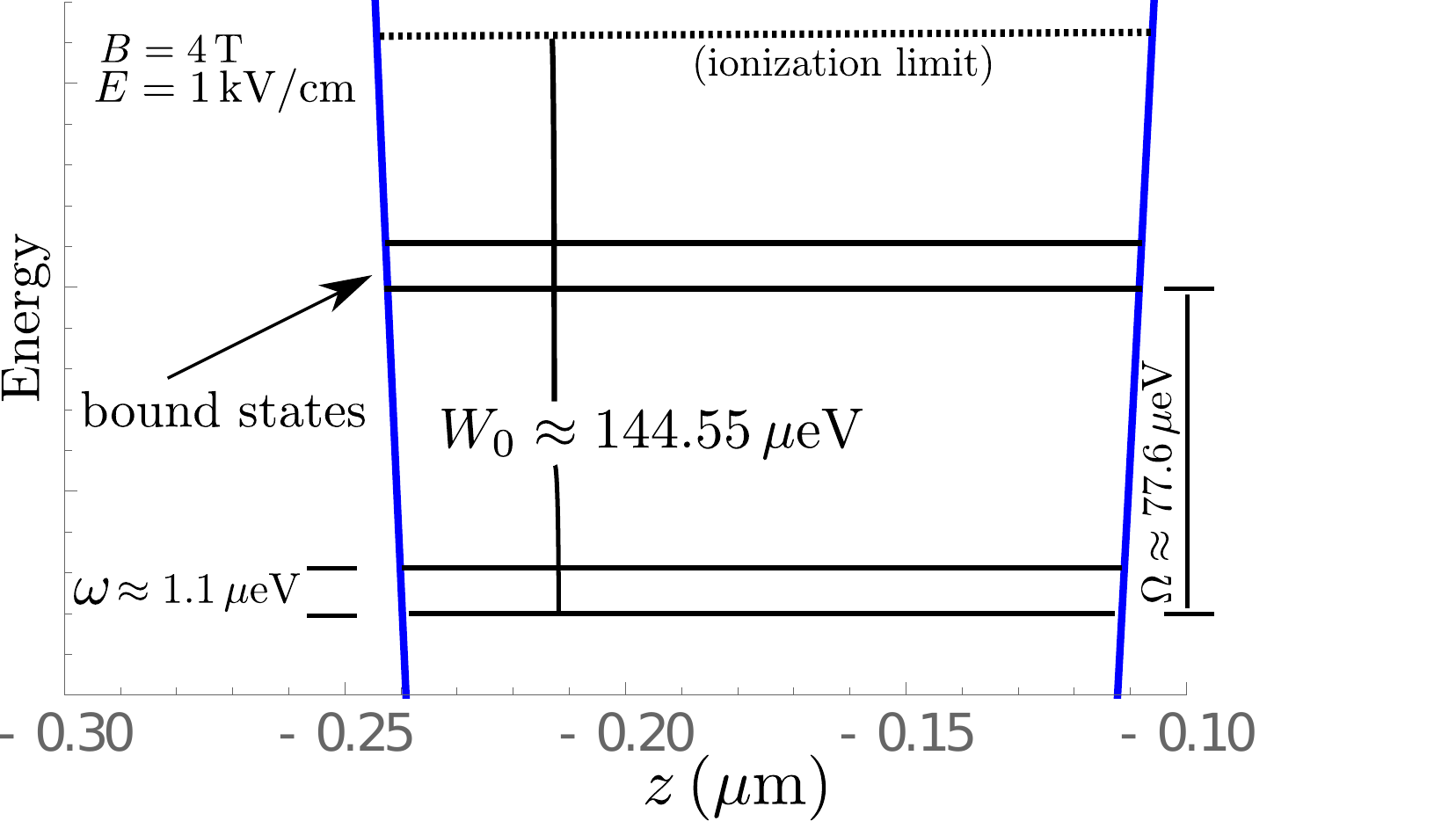}  
\caption{Bound excitonic states in the energetically lowest potential surface 
($x,y=0$). The ground state possesses a binding energy of $144.55\, 
\mu\textrm{eV}$, we find four bound states in total. The external field 
parameter are $B=4\, \textrm{T}$, $E=1\, \textrm{kV/cm}$.}
\label{plot_bound_states}
\end{figure}
In Fig.~\ref{plot_bound_states}, we show the lowest bound excitonic 
giant-dipole states for $B=4\, \textrm{T}$ and $E=1\,\textrm{kV/cm}$, 
respectively. For these field strengths, the ground state possesses an 
approximate binding energy of $W_0\approx144.55\, \mu \textrm{eV}$. In our 
analysis, we estimate the binding energy of a certain state to be the energetic 
separation of the eigenenergy to the ionization limit of the lowest potential 
surface. The binding energies of the excited states are of similar 
order of magnitude, namely in the range between $144\,\mu\textrm{eV}$ and
$68\,\mu\textrm{eV}$. In total, we find four bound states. 

For all applied field strengths, the series of binding energies of 
the giant-dipole states can be cast into the form
\begin{equation}
W_{n,m}=W_0-n\Omega-m\omega, \ n \in \mathbb{N}_0,\ m \in \{0,1 \}, 
\end{equation}
where $\Omega,\omega >0$, $\Omega\gg\omega$, and $W_0$ denotes the 
ground-state binding energy.
\begin{figure}[ht]
\centering
\includegraphics[width=0.45\textwidth]{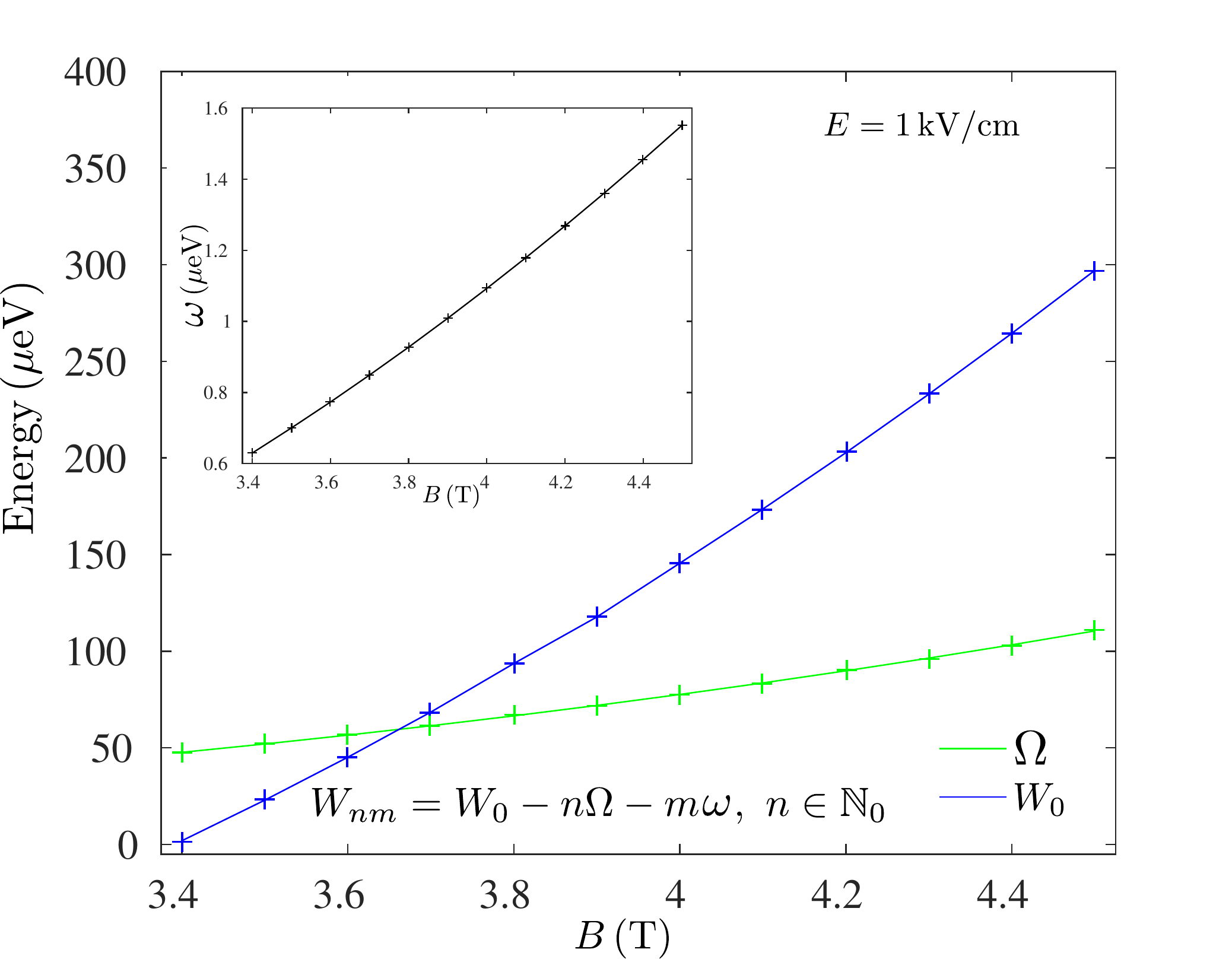}  
\caption{Binding energy $W$ (blue dots) and energy scale $\Omega$ (green 
dots) as a function of the magnetic field $B$. The inset shows the smaller energy scale $\omega$ as a function of $B$.}
\label{energies_2}
\end{figure}
In Fig.~\ref{energies_2}, we show the magnetic-field dependence of the 
binding energy (blue dots) and the larger energy scale $\Omega$ (green 
dots) for $3.4\, \textrm{T} \le B \le 4.5\, \textrm{T}$ for fixed 
electric field strength $E=1\, \textrm{kV/cm}$. One observes that the 
ground-state binding energy increases nearly linearly with increasing magnetic 
field strength from $W_0=1.92\,\mu\textrm{eV}$ 
($B=3.4\, \textrm{T}$) to $W_0=297.05\,\mu\textrm{eV}$ ($B=4.5\,\textrm{T}$). 
The same holds for $\Omega$, which increases from $\Omega=47.67\,\mu\textrm{eV}$ 
($B=3.4\,\textrm{T}$) to $\Omega=110.491\,\mu\textrm{eV}$ ($B=4.5\,\textrm{T}$). 
The inset in Fig.~\ref{energies_2} shows the magnetic-field dependence of the 
smaller energy scale $\omega$, with an almost linear increase from 
$\omega=0.63\,\mu\textrm{eV}$ to $\omega=1.55\,\mu\textrm{eV}$. This can be 
explained by the fact that, with increasing magnetic-field strength, the 
spatial confinement within the potential surface increases as well, which leads 
to larger energetic separation of adjacent energy levels. 

\begin{figure}[h]
\centering
\includegraphics[width=0.45\textwidth]{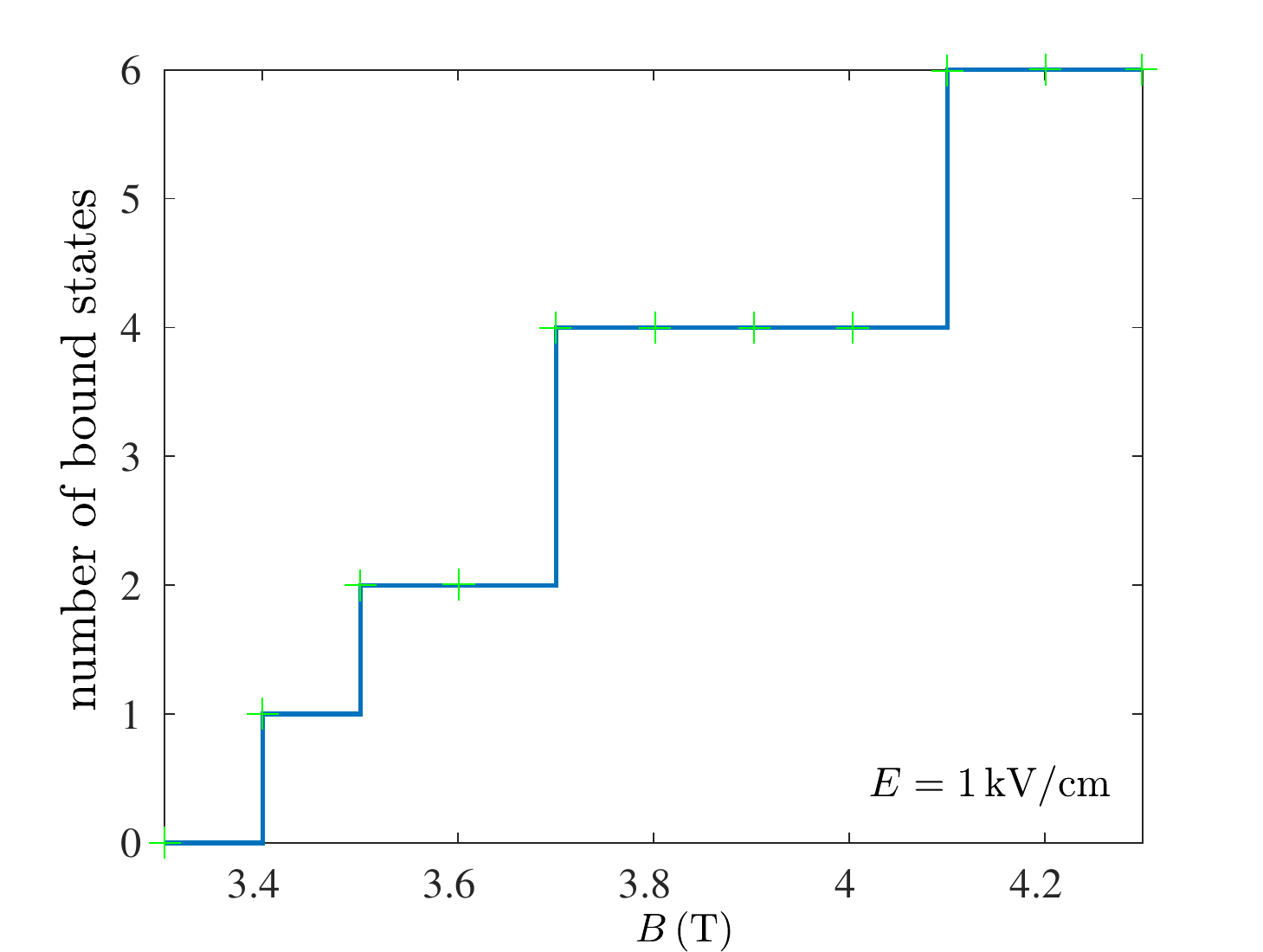}  
\caption{The number of bound excitonic giant-dipole states as a function of the 
magnetic field strength $B$. For $B < 3.4\, \textrm{T}$ no bound states are 
present, from there the number of bound states increase up to six for $B \ge 
4.1\, \textrm{T}$.}
\label{nb_bs}
\end{figure}

The number of bound states depends on whether the large energy scales 
$W_0,\Omega$ and $\omega$ are related to the potential depth of the 
energetically lowest potential surface. In Fig.~\ref{nb_bs}, we show the number of bound states as a function of the 
magnetic field strength for $E=1\,\textrm{kV/cm}$. Below 
$B\lesssim3.4\,\textrm{T}$, the potential well is too shallow and no bound 
excitonic giant-dipole states can be formed. At $B=3.4\,\textrm{T}$, the 
potential becomes slightly deeper than $W_0$, which means that one bound 
state fits into the well. Increasing $B$ leaves the number of bound states 
unchanged as long as the potential well is smaller than $W_0+\omega$ which is 
true for $3.4\,\textrm{T}\le B\le3.5\,\textrm{T}$. Beyond that, we find two 
bound states. This sequence continues until the maximal number of six bound 
states is reached for $B\gtrsim4.3\,\textrm{T}$. Note that one cannot  
arbitrarily increase the magnetic field strength to further increase the number 
of bound states, as the outer potential wells cease to exist when the magnetic 
field contributions are stronger than the Stark term provided by the electric 
field in Eq.~(\ref{Vgd}). 

One effect of the relatively strong magnetic field is that the bound states 
imply an electron-hole separation well below one micrometer, namely 
of around $200\,\textrm{nm}$ for $B=4\,\textrm{T}$. In this case, the 
excitonic dipole moment can be estimated to be around $120\,\textrm{Debye}$, 
which is less than expected from the analysis performed in Ref.~\cite{Kurz2017} 
but still large compared to normal atomic and excitonic dipole strengths.

\begin{figure}[ht]
\centering
\includegraphics[width=0.45\textwidth]{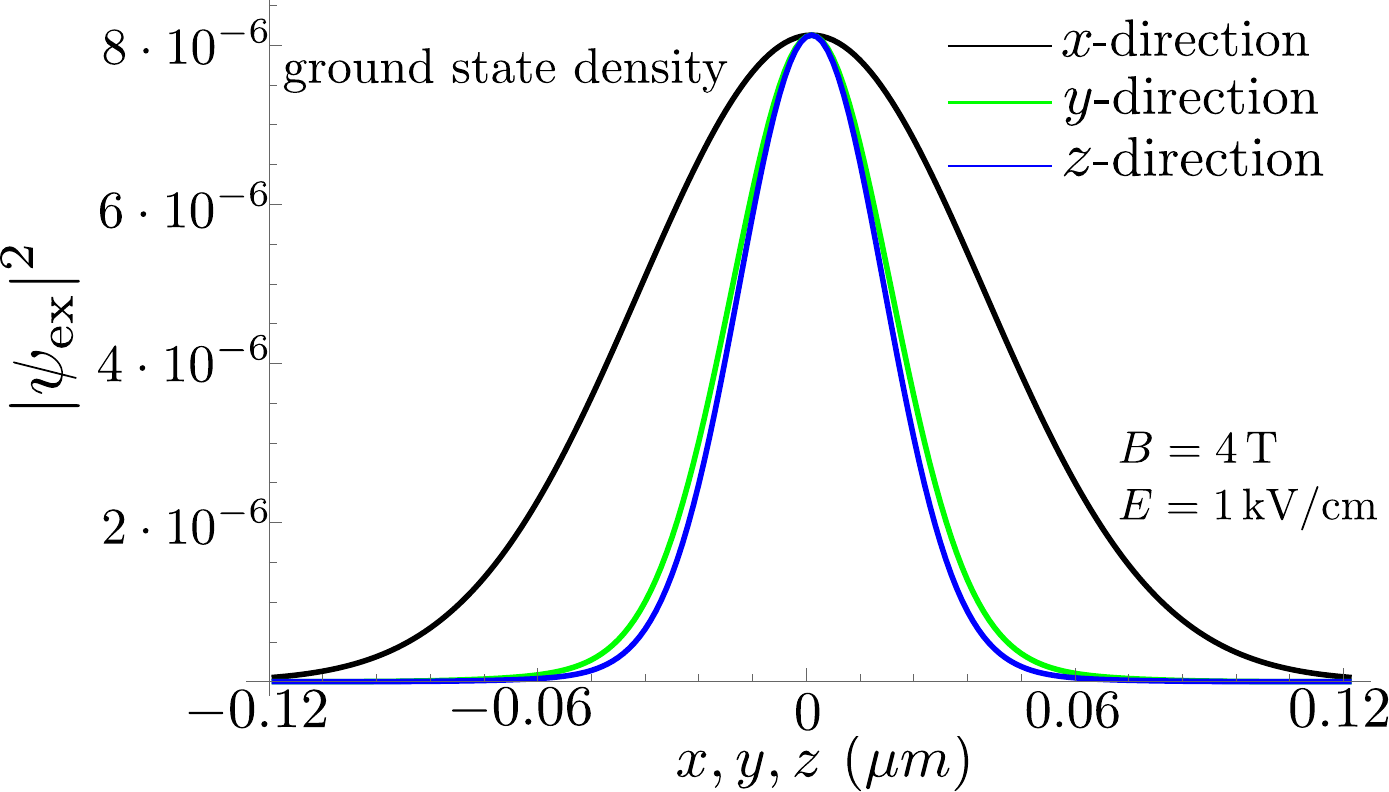}  
\caption{Cuts through the ground-state probability density along the $x$, $y$, 
and $z$-direction, respectively. Due to the different spatial confinement 
($\omega_y/\omega_x \approx 4.4,\ \omega_z/\omega_x \approx 4.1$), the spatial 
extension differs along the different directions.}
\label{wf_gd}
\end{figure}

Apart from the excitonic eigenenergies, the exact diagonalization scheme also
provides the excitonic eigenfunctions. For instance, in Fig.~\ref{wf_gd} 
we show the (Gaussian) ground-state probability density along the $x$, $y$ and 
$z$-directions, respectively. While the spatial extension is nearly equal in 
the $y$ and $z$-directions, the extension in the $x$-direction is much larger.
In particular, for $B=4\,\textrm{T}$ and $E=1\,\textrm{kV/cm}$ we find that 
$\omega_y/\omega_x \approx 4.4$ and $\omega_z/\omega_x \approx 4.1$, a slightly 
smaller spatial confinement in the $y$-direction than in the $z$-direction, 
which is reflected in the ground-state probability density in Fig.~\ref{wf_gd}.
\subsection{Fields in arbitrary directions}
\label{rot_b}

So far, we have analyzed the case of the magnetic and 
electric field being parallel to the $[100]$ and $[001]$ directions, 
respectively. In order to provide some insight into different field 
configurations, we now consider the case that the both magnetic and electric 
fields are oriented along arbitrary directions, whilst still being
perpendicular with respect to one another. Introducing the spherical angles 
$\phi_B$ and $\theta_B$ of the magnetic field vector $\bfmath{B}$, the unit 
vectors $\bfmath{b}$ and $\bfmath{e}$ for the magnetic and electric fields can 
be expressed as
\begin{gather}
\bfmath{b}=
\begin{pmatrix}
\cos(\phi_B)\sin(\theta_B)\\
\sin(\phi_B)\sin(\theta_B)\\
\cos(\theta_B)
\end{pmatrix},\\
\bfmath{e}=
\begin{pmatrix}
-\cos(\phi_B)\cos(\theta_B)\\
-\sin(\phi_B)\cos(\theta_B)\\
\sin(\theta_B)
\end{pmatrix}.
\end{gather}
\begin{figure}[ht]
\centering
\includegraphics[width=0.45\textwidth]{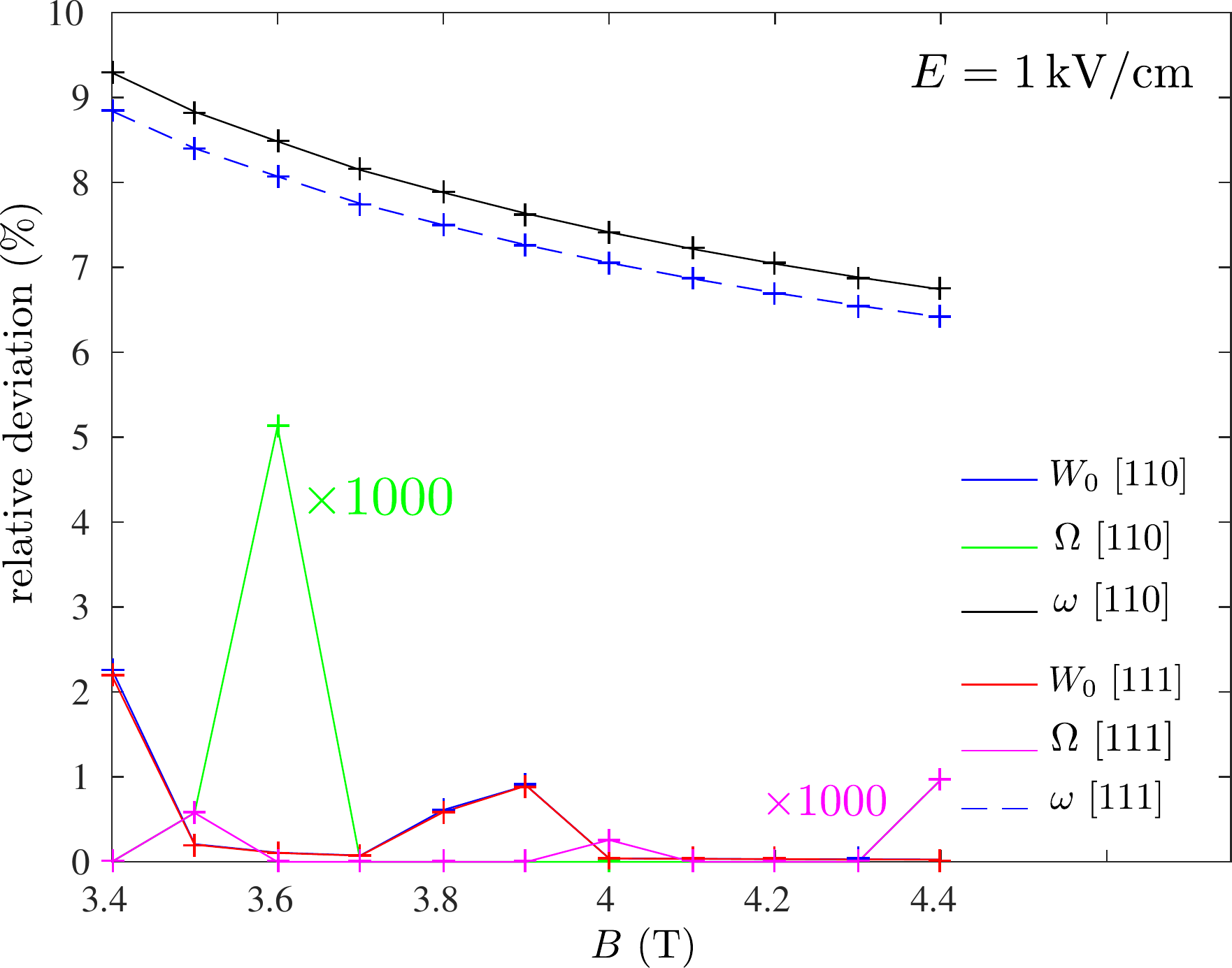}  
\caption{Relative deviation of the energies scales $\Omega, \omega$ and $W_0$ 
for magnetic field orientations $B || [110]$ and $B||[111]$, respectively. The 
largest deviation of around $9\%$ is found for $\omega\ [110]$ and $\omega\ 
[111]$ in the case of $B=0$. For $W_0$ is deviation is smaller, namely $2\%$ and 
less. In the case of $\Omega$ is smallest relative deviations are found with 
$0.005\%$ and much less.}
\label{esd}
\end{figure}
For arbitrary field directions, the potential minimum is still to be found in 
the direction of the electric field. In order to introduce local giant-dipole 
states for the exact diagonalization procedure, one would have to introduce a 
set of local coordinates. However, this inconvenience can be overcome by 
rotating the coordinate system in such a way that the magnetic field direction 
coincides with the quantization axis of the system. In particular, we rotate 
the coordinate system by
\begin{eqnarray*}
R=
\begin{bmatrix}
\cos(\phi_B)\sin(\theta_B) & \sin(\phi_B)\sin(\theta_B) & \cos(\theta_B)\\
-\sin(\phi_B) & \cos(\phi_B) & 0\\
-\cos(\phi_B)\cos(\theta_B) & -\sin(\phi_B)\cos(\theta_B) & \sin(\theta_B)\\
\end{bmatrix},
\end{eqnarray*}
The rotation of the system is performed by rotating the excitonic 
Hamiltonian, in particular, the irreducible tensor representation given by 
Eq.~(\ref{Hex}) transforms by applying Wigner $D$-matrices. In 
App.~\ref{field_dressed_hamiltonian}, we give details of the transformed 
excitonic Hamiltonians for magnetic fields along the $[110]$ and $[111]$ 
directions, respectively.

The calculation of the excitonic giant-dipole eigenenergies for the different 
field configurations has been performed analogously to the magnetic field along 
$[100]$. In particular, the giant-dipole potential surfaces were expanded 
around the potential minimum up to second order, then appropriate basis 
sets were defined in order to perform an exact diagonalization for the 
numerical determination of the eigenenergies. We find that for all 
considered field configurations the excitonic spectra are determined by two 
distinct energy scales as observed in Sec.~\ref{full_spectra}. In 
particular, we find that for all field orientations the energy scales 
$\Omega,\omega$ as well as the ground-state binding energy $W_0$ differ only 
slightly from another. 

In Fig.~\ref{esd}, we show the relative deviation of the energies 
$\Omega,\omega$ and $W_0$ for $B || [110]$ and $B || [111]$ from their values
obtained in the case of $B || [100]$ (see Fig.~\ref{energies_2}). The 
largest deviation is found for $\omega$ with a relative deviations between 
$7\%$ and $9\%$. For increasing magnetic field strength the deviations are 
monotonically decreasing. The relative deviations of $W_0$ the relative are 
even smaller, they are found to be around $2 \%$ for $B=0$ and 
nearly vanish for $B\ge 4\,\textrm{T}$. Furthermore, we see that there are 
hardly any deviations for the energies scales $W_0\, ([110])$ and $W_0\, 
([111])$. Finally, the smallest relative deviations are found for $\Omega\,([110])$ and 
$\Omega\,([111])$, with the largest deviation of merely $0.005\%$ for 
$\Omega\, ([110])$, and around $0.001\%$ for $\Omega\, ([111])$.
\section{Summary and Conclusions}
In the present article, we have calculated the eigenspectra of giant-dipole 
excitons in $\textrm{Cu}_2\textrm{O}$ subject to crossed electric and magnetic 
fields. In particular, we have derived the irreducible tensor representation of 
excitons in crossed electric and magnetic fields in cuprous oxide. In this way, 
the analysis of the excitonic systems for arbitrary field strength and 
arbitrary field configurations is straightforward as the irreducible 
representations can be used to transform the system in such a way that the 
magnetic field coincides with the quantization axis.

In particular, we have calculated the eigenenergies of giant-dipole excitons in 
$\textrm{Cu}_2\textrm{O}$ for arbitrary field strengths and orientations by
applying both an adiabatic approximation as well as an exact diagonalization 
approach. We verify that, in order to find bound excitonic giant-dipole states, 
one requires sufficiently deep potential surfaces. As the depths of the 
considered potential surfaces strongly depend on the applied field strengths, 
bound states are only possible in the limit of weak electric and strong  
magnetic fields. For instance, we find bound states for field strengths of 
around $E\approx 1\,\textrm{kV/cm}$ and $3.4\,\textrm{T}\le 
B\le4.3\,\textrm{T}$. For all field orientations, the corresponding level 
spacings are determined by two energy scales which are of the order of 
$1-2\,\mu\textrm{eV}$ and $100\,\mu\textrm{eV}$, respectively. The number of 
bound states is comparable small, for the considered field strengths we find 
between one and six bound states for all field orientations.

An open question is the experimental preparation and verification of the 
the existence of these excitonic giant-dipole states. The latter could, 
in principle, be achieved via spectroscopic measurements of the excitonic 
resonances which should be visible in microwave spectroscopy. An 
alternative approach might be the direct measurement of the large electric 
dipole moment which can be estimated to be of the order of several tens of
thousand Debye. 

Another yet unsolved question is how to prepare those exotic excitonic states. 
Due to the large spatial separation between the outer potential wells and the 
Coulomb-dominated region, a direct radiative transfer via external lasers is 
unlikely as the overlap between the giant-dipole wave functions and 
low-lying exciton states in the inner region is very small. 

However, one possible approach may be to use the field-free excitation of 
highly excited Rydberg excitons. Applying time-dependent external fields 
hereafter, one might be able to adiabatically transfer the Rydberg state into 
the desired field-dressed giant-dipole configuration. For the determination of a 
possible propagation scheme one has to consider that, although classical 
trajectory simulations has already provided some understanding for a possible 
preparation scheme for atomic giant-dipole states \cite{Averbukh1999}, the setup 
for excitonic states is more complicated due to the complex spin-structure. In 
particular, one has to consider six distinct coupled potential surfaces, 
causing non-adiabatic state transfer among those. In order to include 
non-adiabatic transitions between the potential surfaces, one may adapt a method 
from molecular dynamics calculations known as surface hopping 
\cite{Herman1984,Tully1990} which partially incorporates the non-adiabatic 
effects by including excited adiabatic surfaces in the calculations, and 
allowing for transitions between these surfaces. An alternative approach might 
be to perform a full quantum mechanical analysis to achieve an optimal state 
transfer starting from an appropriate initial excitonic state. This belongs to 
a general class of problems known as control theory 
\cite{Werschnik2007,Raesaenen2007}, where one is interested in finding a 
protocol to change addressable system parameters such that a certain 
optimal criterion is achieved. 

Yet another possibility to create excitonic giant-dipole states might be to 
directly start in the field-dressed configuration and to excite ground-state 
excitons directly into the continuum, that may recombine into states localized 
in the outer potential wells due to radiative decay, interspecies scattering 
events or phonon-induced de-excitation. Especially the last decay channel might 
be of particular interest as it is induced by the solid-state environment and 
which is not present in ultra-cold atomic gases.  
In summary, the preparation of excitonic giant-dipole states provides a 
plethora of interesting research directions that can be addressed in future 
theoretical as well as experimental studies.

\begin{acknowledgments}
We acknowledge support by the DFG SPP 1929 GiRyd funded by the 
Deutsche Forschungsgemeinschaft (DFG).
\end{acknowledgments}
\appendix

\section{Excitonic parameters}
\label{app_exc_paramteter}
In Tab.~\ref{table} we list the excitonic parameters used throughout this 
work.
\begin{table}[hbt]
\centering
\begin{tabular}{ c c c }
  \hline			
Luttinger parameters & $\gamma_1,\gamma_2,\gamma_3$ & $1.818,0.803,-0.397$\\
& $ \gamma^{'}_{1},\kappa$ & $2.83,-0.5$  \\ 
Hartree energy & $\mathcal{H}_{\rm ex}$ &$170\, \rm meV$ \\
(excitonic) Bohr radius & $a_{\rm ex}$ & $1.15\, \rm nm$\\
  magnetic flux density & $B_{\rm ex}$ & $520.6\, \rm T$ \\
  electric field strength & $E_{\rm ex}$ & $1.518\, \rm MV/cm$ \\
  momentum & $P_{\rm ex}$ & $4.8 \times 10^{-2}\hbar/a_0$ \\
  gap energy & $E_g$ & $2.17208\, \rm eV$\\
  spin-orbit coupling &$\Delta$ & $133.8\, \rm meV$\\
Bohr magneton &$\mu_B$& $57.88\, \mu \rm eV/T$\\
  \hline  
\end{tabular}
\caption{Excitonic Hartree energy $\mathcal{H}_{\rm ex}$, Bohr radius $a_{\rm 
ex}$, external field strengths $(B_{\rm ex},E_{\rm ex})$ and momentum $P_{\rm 
ex}$. In addition, the spin-orbit and magnetic coupling $(\Delta,\mu_B)$ is 
presented as well as the Luttinger parameters used throughout this work.}
\label{table}
\end{table}
\section{Williamson's theorem} 
\label{app_williamson}
Let $M$ be a positive-definite symmetric real $2n \times 2n$ matrix. In this case the following theorem holds \cite{Wiliamson1936}:
\begin{itemize}
 \item [(i)] There exists $S \in \textrm{Sp}(2n,\mathbb{R})$ such that
 \begin{eqnarray*}
 M&=&S^{T}DS,\ \ \ D=\textrm{diag}(\Lambda,\Lambda),\\ 
\Lambda&=&\textrm{diag}(\lambda_1,...,\lambda_n),\ \ \ \textrm{with}\ \ \  
 \lambda_i \in \mathbb{R}^{>0}.
 \end{eqnarray*}
\item[(ii)]The entries $\lambda_i$ of $\Lambda$ are defined by the condition 
that $\pm i \lambda_i$ is an eigenvalue of $J_n M$ where
\begin{equation*}
J_n= 
\begin{bmatrix}
0 & 1_{n}\\ 
-1_{n} & 0\\ 
\end{bmatrix}. 
\end{equation*}
\item[(iii)] The sequence $\lambda_1,...,\lambda_n$ does not depend, up to a 
reordering of its terms, on the choice of $S$ diagonalizing $M$.
\end{itemize}
We introduce ${\bf q}=S{\bf x}$ with $S \in \textrm{Sp}(6,\mathbb{R})$. In this case, the 
canonical commutator relations are preserved, i.e.\
$[q_i,q_j]=iJ_{3,ij}$.
\section{Irreducible representation of kinetic and potential energy 
\label{t_v_irr_rep}}
The irreducible tensor representation of the kinetic energy term 
$T$ and the potential energy $V$ for a magnetic field orientation of 
$\bfmath{B}||[100]$ discussed in Sec.~\ref{gd_Hamiltonian} is given by
\begin{eqnarray*}
&\ &T=\frac{\bfmath{\pi}^2}{2}- \frac{\mu'}{3}\{ (\Pi^{(2)} \cdot I^{(2)} ) \}\\
&\ &+\frac{\delta'}{3} \bigg( \sum_{k= \pm 4} \{ [\Pi^{(2)} \times I^{(2)}]^{(4)}_{k} \}
+\frac{\sqrt{70}}{5} \{ [\Pi^{(2)} \times I^{2}]^{(4)}_{0} \} \bigg),
\end{eqnarray*}
\begin{eqnarray*}
V&=&\left(\Omega_1 \tilde{K}^2+E^{(1)} \cdot r^{(1)}-\frac{1}{r}\right)- \frac{\lambda'}{3} (\tilde{K}^{(2)} \cdot I^{(2)} )\\
&\ &+\frac{\xi'}{3} \left( \sum_{k= \pm 4}[\tilde{K}^{(2)} \times I^{(2)}]^{(4)}_{k}+\frac{\sqrt{70}}{5}  [\tilde{K}^{(2)} \times I^{2}]^{(4)}_{0}  \right)\\
&\ &+H_{\rm so}+H_{\rm B}
\end{eqnarray*}
with $\lambda'=\frac{2}{5}(\Omega_3+ \Omega_2),\ \xi'=\Omega_3/3-\Omega_2/2$ and
\begin{eqnarray*}
H_{\rm so}&=&\frac{2}{3}\bar{\Delta}\left( 1 + S^{(1)}_{h}\cdot I^{(1)} \right),\\ H_{\rm B}&=&\bar{\mu}_{B}[(3 \kappa +\frac{g_s}{2})I^{(1)}\cdot B^{(1)}-g_s S^{(1)}_{h}\cdot B^{(1)}].
\end{eqnarray*}
\section{The excitonic Hamiltonian in adiabatic approximation}
\label{app_gd_ham_williamson}
In Eq.~(\ref{H_eff_w}), we introduced the matrix representation of the 
excitonic Hamiltonian,
\begin{eqnarray*}
H^{(\alpha)}_{\rm eff}= \bfmath{x}^T \mathcal{H}^{(\alpha)}_{\rm eff} \bfmath{x},\ \ \ \textrm{with}\ \ \ 
\mathcal{H}^{(\alpha)}_{\rm eff}&=&A^{T}MA,\\
\bfmath{x}&=&(x\ y\ z\ p_x\ p_y\ p_z)^T.
\end{eqnarray*}
The matrices
\begin{eqnarray*}
M=\textrm{diag}(P,Q)\ \ \ \textrm{and}\ \ \ 
A=
\begin{bmatrix}
I_{3} & 0 \\
W& I_{3}\\
\end{bmatrix}
\end{eqnarray*}
are given in block form, the submatrices $P,Q,W$ and $I_{3}$ are given by
\begin{eqnarray*}
P&=&
\frac{1}{2}
\begin{bmatrix}
C^{(\alpha)}_{xx} & C^{(\alpha)}_{xy} &C^{(\alpha)}_{xz}\\
C^{(\alpha)}_{xy}& C^{(\alpha)}_{yy}& C^{(\alpha)}_{yz}\\
C^{(\alpha)}_{xz} & C^{(\alpha)}_{yz} & C^{(\alpha)}_{zz}\\
\end{bmatrix},\ 
Q=
\begin{bmatrix}
\frac{1}{2 \mu^{(\alpha)}_{x}} & \gamma^{(\alpha)}_{xy} & \gamma^{(\alpha)}_{xz} \\
\gamma^{(\alpha)}_{xy} & \frac{1}{2 \mu^{(\alpha)}_{y}} & \gamma^{(\alpha)}_{yz}\\
\gamma^{(\alpha)}_{xz}&\gamma^{(\alpha)}_{yz} & \frac{1}{2 \mu^{(\alpha)}_{z}}\\
\end{bmatrix}\\
W&=&
\frac{1}{2}
\begin{bmatrix}
0 & 0 &0\\
0& 0& q\tilde{B}^{(\alpha)}_{z}\\
0 & -q\tilde{B}^{(\alpha)}_{z} & 0\\
\end{bmatrix},\
I_{3}=
\begin{bmatrix}
1 & 0 &0\\
0& 1& 0\\
0 & 0 & 1\\
\end{bmatrix}.\\
\end{eqnarray*}
\section{Field-dressed excitonic Hamiltonians 
\label{field_dressed_hamiltonian}}
The irreducible tensor representations of field-dressed excitonic 
Hamiltonians for $\bfmath{B}||[110]$ and $\bfmath{B}||[111]$, 
respectively, are listed below.
\subsection*{Magnetic field in [110] direction}
\begin{eqnarray*}
&\ &\hspace{-0.55cm}H_{\rm ex}= \frac{\bfmath{\pi}^2}{2}- \frac{\mu'}{3}\{ (\Xi^{(2)} \cdot I^{(2)} )
+\frac{\delta^{\prime}}{4}\big(\sum_{k=\pm4}\{[\Xi^{(2)}\times I^{(2)}]^{(4)}_{k}\} \big)\\
&\ &\hspace{-0.5cm}-\frac{\sqrt{7}}{6}\delta^{\prime} \big( \sum_{k=\pm 2}\{[\Xi^{(2)}\times I^{(2)}]^{(4)}_{k}\}+\sqrt{\frac{1}{10}}\{[\Xi^{(2)}\times I^{(2)}]^{(4)}_{0} \} \big)\\
&\ &+\bigg(\Omega_1 \tilde{K}^2+E^{(1)} \cdot r^{(1)}-\frac{1}{r}\bigg)+H_{\rm so}+H_{\rm B}.
\end{eqnarray*}
\subsection*{Magnetic field in [111] direction}
\begin{eqnarray*}
&\ &H_{\rm ex}=\frac{\bfmath{\pi}^2}{2}- \frac{\mu'}{3}\{\Xi^{(2)} \cdot I^{(2)} )\}\\
&\ &\hspace{-0.475cm}+\frac{4}{27}\delta^{\prime}  \big(\sum_{k= \pm 3}k\{ [\Xi^{(2)} \times I^{(2)}]^{(4)}_{k} \}
-\sqrt{\frac{63}{10}}\{[\Xi^{(2)}\times I^{(2)}]^{(4)}_{0}\} \big) \\
&\ &+\left(\Omega_1 \tilde{K}^2+E^{(1)} \cdot r^{(1)}-\frac{1}{r}\right)+H_{\rm so}+H_{\rm B}.
\end{eqnarray*}
\end{document}